\documentclass[final,3p,times,authoryear]{elsarticle}

\usepackage{amssymb}
\usepackage{amsmath,amssymb,amsfonts}
\usepackage[caption=false]{subfig}
\newcommand{\del}{\partial}
\newcommand{\bs}{\boldsymbol}
\newcommand{\rmd}{\mathrm{d}}
\newcommand{\rmi}{\mathrm{i}}

\newcommand{\rmin}{\mathrm{in}}
\newcommand{\rmout}{\mathrm{out}}

\newcommand{\rmsh}{\mathrm{sh}}

\begin{document}

\begin{frontmatter}



\title{Analytical solution for a vibrating rigid sphere with an elastic shell in an infinite linear elastic medium}


\author[inst1]{Evert Klaseboer}
\affiliation[inst1]{organization={Institute of High Performance Computing},
            addressline={1 Fusionopolis Way},
            city={Singapore},
            postcode={138632}, 
            country={Singapore}}
            
\author[inst2]{Qiang Sun\corref{cor1}}
\cortext[cor1]{qiang.sun@rmit.edu.au}
\affiliation[inst2]{organization={Australian Research Council Centre of Excellence for Nanoscale BioPhotonics, School of Science, RMIT University},
            city={Melbourne},
            postcode={VIC 3001}, 
            country={Australia}}

\begin{abstract}
The analytical solution is given for a vibrating rigid core sphere, oscillating up and down without volume change, situated at the center of an elastic material spherical shell, which in turn is situated inside an infinite (possible different) elastic medium. The solution is based on symmetry considerations and the continuity of the displacement both at the core and the shell - outer medium boundaries as well as the continuity of the stress at the outer edge of the shell. Furthermore, a separation into longitudinal and transverse waves is used. Analysis of the solution shows that a surprisingly complex range of physical phenomena can be observed when the frequency is changed while keeping the material parameters the same, especially when compared to the case of a core without any shell. With a careful choice of materials, shell thickness and vibration frequency, it is possible to filter out most of the longitudinal waves and generate pure tangential waves in the infinite domain (and vice-versa, we can filter out the tangential waves and generate longitudinal waves). When the solution is applied to different frequencies and with the help of a fast Fourier transform (FFT), a pulsed vibration is shown to exhibit the separation of the longitudinal (L) and transverse (T) waves (often called P- and S-waves in earthquake terminology). 
\end{abstract}


\begin{highlights}
\item Adding a shell profoundly changes the dynamics of the system
\item Separation of T and L waves through a FFT transform
\item Analytical solution can describe complex phenomena
\item Through a careful choice of materials, shell thickness and frequency, we can either filter out transverse or longitudinal waves. 
\end{highlights}


\end{frontmatter}


\section{Introduction}

Dynamic linear elastic problems appear on many length scales. On the large scales we can find problems relating to earthquakes and other geophysics problems.
On the small scales ($\mu$m to mm - scale), we can think of the application of various biomedical treatments with ultrasound or shockwaves, where the biomaterial is often regarded as simply acoustic, even though this assumption might not always be justified~\citep{Rapet2019}. Advances in the development of ultrasonics and microfluidics
have also renewed the interest in this area~\citep{Dual2012}, such as cell trapping as well as ultrasonic inspection. With the availability of MHz acoustic transducers in recent years, applications in these areas are probably going to increase.  At intermediate length scales, say 1 m, one can think of studies regarding sound and vibration reduction caused by moving parts of machinery. 

Although numerical methods can be employed to solve dynamic linear elastic problems, they do not necessary give physical insight which can only be obtained from analytical solutions~\citep[Chap 3]{hills2021}.

Steady state analytical solutions are rare but do exist~\citep{Lim2006}. Analytical solutions for dynamic linear elastic problems are even rarer (or at least not very well known). Sneddon and Berry on page 126 wrote ``There are very few exact solutions even of these steady state equations and such as they are limited to spheres and cylinders''. Moreover, those analytical solutions are not only rare, but also they often do not show the `full' range of physics of real systems; that is, some severe limitations are being imposed on the solution, such as the solution for a radially oscillating sphere which only gives longitudinal waves without transverse waves~\citep{Grasso2012}. Solutions that show both longitudinal and transverse waves do exist though, for example the elastic scattering wave solution by~\cite{Hinders1991} which is very similar to the Mie theory~\citep{Mie1908} in electromagnetics. A drawback of this solution is that an infinite sum of Bessel functions is needed to calculate the solution. Although elegant, it is not easy to guess how many terms one needs in this sum when using Bessel functions (the accuracy can even go down again, when too many terms are included). Classical authors such as Lamb and Love~\citep{Love1892} already described various analytical solutions for example for the internal resonances for elastic spheres or the solution for the elastic material in between two concentric spheres. \cite{Papargyri2009} presented some analytical solutions for gradient elastic solids.

\cite{KlaseboerJElast2019} presented an analytical solution for a rigid sphere vibrating in an infinite elastic medium. The current article can be considered as a significant extension of this theory, by adding an elastic shell to the rigid core. This `shell' solution turns out to be much richer in physical phenomena than the case without a shell and is the focus of the current work. The solution is free of infinite sums and is relatively easy to calculate and visualize (a clear advantage that the 19th century classical authors did not have), which makes it ideal to validate numerical tools. 

\begin{figure*}[t]  
\begin{center} 
\includegraphics[width=0.75\textwidth]{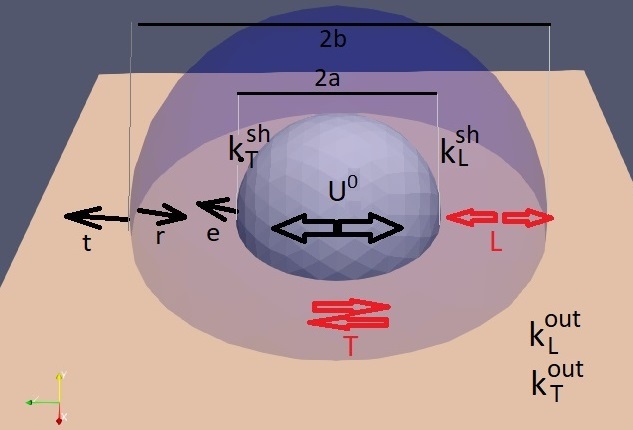}
    \caption{Schematic illustration of the problem under consideration; a rigid core sphere with radius $a$ oscillates periodically with an amplitude $U^0$ with frequency $\omega$ and is surrounded by a concentric spherical shell with radius $b$ and material constants $k_L^{\rmsh}$ and $k_T^{\rmsh}$. Both spheres are embedded in an infinite material with material constants $k_L^{\rmout}$ and $k_T^{\rmout}$, which may or may not be identical to the shell constants. Elastic waves are being emitted (labeled `e'), reflected (labeled `r') and transmitted (labeled `t') towards infinity. Both longitudinal (L) waves and transverse (T) waves can exist in this system. } \label{Fig:Explanation}
\end{center}
\end{figure*}

In Fig~\ref{Fig:Explanation}, an illustration of the problem is given, the core rigid sphere with radius $a$ oscillates with an amplitude $U^0$ and angular frequency $\omega$ inside a concentric spherical shell with radius $b$ and density $\rho_{\rmsh}$. In doing this emitted (`e') or outgoing waves are generated in the shell. Reflected waves (`r') can also occur. Finally in the external infinite domain with density $\rho_{\rmout}$ transmitted (`t') waves can be generated. The material constants in the shell and the infinite domain are not necessarily the same. 
As we will see in the latter part of this work, the solution for a rigid vibrating core with a shell around it shows some surprising physics (resonance peaks for T and/or L waves when the vibrating frequency is changed), which do not occur for a core without such a shell, where a very smooth spectrum is obtained.

Also, the analytical solutions with different parameters can be used as non-trivial test cases for numerical methods, for instance, one example has been given for validating a boundary element method in \ref{App:BEM}. Furthermore, the vibrating spherical core-shell system could possibly be used as a simple and elegant template for practical applications, such as, to design spherical piezoelectrical actuators to emit or harvest energy which efficiency highly depends on the material properties and frequency response~\citep{Covaci2020}, possibly with multiple spherical core-shell structures placed in arrays. Another way to generate either longitudinal or transverse waves can be achieved by vibrating the metallic core sphere using magnetic means, which can make the spherical core-shell system become a heat generator when it converts magnetic energy to heat via relaxation processes and hysteresis losses~\citep{Schmidt2007}. Along this line, our analytical solution can be used to optimize the design of hyperthermia agents using magnetic beads for cancer treatments~\citep{Philippova2011}. There are likely other applications in which a spherical core-shell system is a good approximation of a real physical system.

The structure of this work is organized as follows. In Sec.~\ref{sec:anasol}, we demonstrate the derivation of the analytical solution for a vibrating rigid core with a shell in an infinite elastic medium and the detailed steps are given in~\ref{App:AppA} and~\ref{App:AppB}. In Sec.~\ref{sec:results}, we study the elastic wave phenomena at different oscillation frequencies followed by some discussions in Sec.~\ref{sec:discussion} which shows the limit case of solution and pulsed time domain solutions using the fast Fourier transform. The conclusion is given in Sec.~\ref{sec:conclusion}.

\section{Dynamic elastic waves} \label{sec:anasol}
\subsection{General theory}

Within the approximation of small deformations and small stresses, the Navier equation for dynamic linear elasticity in the frequency domain can be written as~\citep{KlaseboerJElast2019,Love1892,Pelissier2007}
\begin{align}  \label{eq:Navier}
   c^2_L\nabla (\nabla \cdot \boldsymbol{u}) - c^2_T \nabla \times \nabla \times \boldsymbol{u} + \omega^2 \boldsymbol{u} = \boldsymbol{0}
\end{align}
where $\boldsymbol{u}$ is the (complex valued) displacement vector, $\omega$ is the angular frequency, and the constants $c_L$ and $c_T$ are the longitudinal dilatation and transverse shear wave velocities, respectively, that are defined in terms of the Lam\'e constants $\lambda$, $\mu$ and the density $\rho$~\citep{LandauLifshitz,Bedford1994}:
\begin{equation} \label{eq:cTcL}
\begin{aligned}  
  c^2_L = (\lambda + 2 \mu)/\rho\quad ; \quad
  c^2_T = \mu/\rho. 
\end{aligned}
\end{equation}
Eq.~(\ref{eq:Navier}) essentially expresses the equilibrium of the elastic forces (the first two terms) and the inertial forces (the third term)\footnote{Here we have ignored volume forces and thermoelastic effects~\citep{Ruimi2012}}. It is well known that the displacement $\boldsymbol{u}$ can be decomposed into a transverse part $\boldsymbol{u}_{T}$ and a longitudinal part $\boldsymbol{u}_L$ as:
\begin{align}  \label{eq:HelmDec}
   \boldsymbol{u} = \boldsymbol{u}_{L}+\boldsymbol{u}_{T},
\end{align}
with $\bs{u}_T$ being divergence free and $\bs{u}_L$ being curl free, thus: 
\begin{align}\label{eq:divCurlZero} 
    \boldsymbol{\nabla} \boldsymbol{\cdot} \boldsymbol{u}_{T} = 0 \quad ; \quad
    \boldsymbol{\nabla} \times \boldsymbol{u}_{L} = \boldsymbol{0}.
\end{align}
In this work, we will refer to the longitudinal waves as ``L" and to the transverse waves as ``T". These two sorts of waves are also often referred to as pressure waves and shear waves, respectively. Introducing Eq. (\ref{eq:HelmDec}) into Eq. (\ref{eq:Navier}) and considering the relations in Eq.~(\ref{eq:divCurlZero}), we obtain
\begin{align} \label{eq:HelmuTuL} 
  \nabla^2 \boldsymbol{u}_T + k^2_T \boldsymbol{u}_T = \boldsymbol{0} \quad ;\quad
  \nabla^2 \boldsymbol{u}_L + k^2_L \boldsymbol{u}_L = \boldsymbol{0},
\end{align}
where $k_T = \omega/ c_T$ and $k_L = \omega/ c_L$ are the transverse and longitudinal wavenumbers, respectively. Thus both the transverse displacement $\bs{u}_T$ and the longitudinal displacement $\bs{u}_L$ satisfy the Helmholtz equation, yet with different wavenumbers. As is obvious from Eq.~(\ref{eq:cTcL}), the longitudinal wave velocity is always greater than the transverse wave speed, thus $c_L^2 \ge 2 c_T^2$ or in terms of wavenumbers $k_T \ge \sqrt{2}k_L$.

\subsection{Theory for vibrating spheres}\label{sec:coreshellsol}
As shown in Fig.~\ref{Fig:Explanation}, we impose that the geometry under consideration consists of a rigid core with radius $r=a$, surrounded by another concentric sphere with radius $r=b$. The material in between the two spheres (the shell) is elastic and indicated with `sh'. The core-shell sphere combination is situated inside a different external outer elastic material referred to as `out'. Since the core only vibrates along the $z$-axis, due to symmetry, we look for a solution that has a zero azimuthal $\varphi$ component for both the displacement and the stress (a similar framework was used by the authors to calculate the acoustic boundary layer around a vibrating sphere, see \cite{KLaseboerPhFl2020} \footnote{\cite{KLaseboerPhFl2020} studied acoustic boundary layers around a sphere in fluid dynamics. The same governing equations  appear  as  in  elasticity,  except  for  the  difference  that $k_L$ and $k_T$ are  now  complex  numbers  and that ‘$\bs u$’ is now the velocity instead of the displacement. The focus was there on the phenomenon of ‘streaming’ which is a second order effect which causes a slow mean flow on top of the flow caused by the oscillation of the sphere. This non linear effect does not appear here. }). Such a solution can be written as:
\begin{equation} \label{eq:phi_h}
\begin{aligned}
    \boldsymbol u = \bs{u}_L + \bs{u}_T&= \nabla \left((\boldsymbol x \cdot \boldsymbol u^0) \frac{\phi(r)}{r}\right) + \nabla \times \left( (\boldsymbol x \times \boldsymbol u^0) \frac{h(r)}{r}\right)\\
    &= \nabla [\phi(r) \cos{\theta}] U^0 \quad \; - \nabla \times [h(r) \sin{\theta}  \bs{e}_{\varphi}] U^0
    \end{aligned}
\end{equation}
where we have used spherical coordinates $(r,\theta,\varphi)$, $\bs{e}_{\varphi}$ is the unit vector in the $\varphi$ direction and $\bs{x}$ is the position vector $\bs{x}=(x,y,z)$. Two radial functions, $h(r)$ and $\phi(r)$, are to be determined\footnote{The solution with $\phi$ and $h$ can only present solutions in the plane made by the vectors $\boldsymbol x$ and $\boldsymbol u^0$. There are other analytical solutions for a spherical configuration that cannot be described with the $h -\phi$ framework. For example for a sphere periodically rotating back and forth with frequency $\bs{\Omega}$ around the $z$-axis, the following analytical solution can be found
    $\boldsymbol u = \boldsymbol u_T = -\frac{a^3}{e^{\rmi k_Ta}} \frac{1}{(\rmi k_Ta-1)} \nabla \times \left[\frac{e^{\rmi k_Tr}}{r} \boldsymbol \Omega \right]$.
The solution now only consists of a transverse part, while there is no longitudinal component. } in which the term $\phi\equiv \phi(r)$ is a potential function and the term $h\equiv h(r)$ is inspired by the $h$-function in electrophoresis problems~\citep{Jayaraman2019, Ohsima1983}. It can easily be seen that the term with $\phi$ corresponds to the curl free vector $\bs{u}_L$ and the term with $h$ to the divergence free vector $\bs{u}_T$ (remember that $\nabla \times \nabla = \bs{0}$ and $\nabla \cdot \nabla \times = 0$). A constant vector $\bs{u}^0$ is introduced with length $|\bs{u}^0|=U^0$. It represents the amplitude of the displacement of the core sphere in the frequency domain. For the time being we will take $\bs{u}^0 =(u_x,u_y,u_z) = (0,0,U^0)$, which means that in the time domain this vector oscillates as $(0,0,U^0\cos(\omega t))$. 

Analytical solutions exist, for example a sphere harmonically changing its volume has an analytical solution\footnote{For a radially volume changing sphere the solution for the displacement is:
    $\boldsymbol u = \boldsymbol u_L = \frac{e^{\rmi k_Lr}}{r^3}(\rmi k_L r -1) \boldsymbol x$
}, yet this solution only shows L-waves and does not have any T-waves. It is therefore desirable to have some analytical solutions that at least show both L and T waves simultaneously. Eq.~(\ref{eq:phi_h}) will turn out to be sufficient to describe the displacement field caused by the vibration of a rigid core sphere, surrounded by an elastic shell, situated in an infinite other elastic material. The function $\phi$ is related to the Helmholtz equation as $\nabla^2 (\phi \boldsymbol x/r) + k_L^2 (\phi \boldsymbol x/r)= \boldsymbol 0$, while $h$ satisfies $\nabla^2 (h \boldsymbol x/r) + k_T^2 (h \boldsymbol x/r) =\boldsymbol 0$. This essentially implies that both $\phi(r) \cos (\theta)$ and $h(r) \cos (\theta)$ satisfy the Helmholtz equation.

The task at hand is now to determine the two functions $\phi$ and $h$. Since the material properties of the shell and the outer material are different, we will search for $\phi^{\rmsh}$ and $h^{\rmsh}$ for the shell solution and $\phi^{\rmout}$ and $h^{\rmout}$ for the external domain. We will describe two different paths, one using tensor notation and another approach for readers more familiar with spherical coordinate systems and Bessel functions. Both approaches of course will lead to the same answer. The problem we wish to solve is to get the displacement field caused by the motion $\bs{u}^0$. In order to do so, we need to satisfy that both displacements and stresses are continuous across boundaries, that is: there are no gaps or stress jumps in the material boundaries. Thus the displacement at $r=a$ must obey $\bs{u}=\bs{u}^0$ and at $r=b$, both $\bs{u}$ and the traction $\bs{f}$ must be continuous. 

Eq.~(\ref{eq:phi_h}) can be written in an alternative more convenient way (note that we have deliberately kept the terms $h/r$ and $\phi/r$) by separating the terms with $\bs{u}^0$ and $\bs{x}$. For the shell we get:
\begin{equation}
\begin{aligned} \label{eq:u_in}
    \bs{u}^{\rmsh} = \left[ -r \frac{\rmd}{\rmd r} \left(\frac{h^{\rmsh}}{r}\right) - 2 \frac{h^{\rmsh}}{r} + \frac{\phi^{\rmsh}}{r}\right]\bs{u}^0 + \frac{\rmd}{\rmd r}\left(\frac{h^{\rmsh}}{r} +\frac{\phi^{\rmsh}}{r} \right) \frac{\bs{x} \cdot \bs{u}^0}{r} \bs{x}
    \end{aligned}
\end{equation}
where the shell solution consist of an `expanding' (subscript `e') and a `reflected' (subscript `r') wave  with 
\begin{equation}
\begin{aligned}
h^{\rmsh}(r)=h_e(r) + h_r(r)=-a C_e^T \exp(\rmi k_T^{\rmsh} r) G(k_T^{\rmsh} r) -a C_r^T \exp(-\rmi k_T^{\rmsh} r) G^*(k_T^{\rmsh} r), \\  \phi^{\rmsh}(r)=\phi_e(r) + \phi_r(r)=-a C_e^L \exp(\rmi k_L^{\rmsh} r) G(k_L^{\rmsh} r) -a C_r^L \exp(-\rmi k_L^{\rmsh} r) G^*(k_L^{\rmsh} r)
\end{aligned}
\end{equation}
with $G^*$ the complex conjugate of $G$ (also note the `-' sign in the $C_r$ exponentials)\footnote{Note that in terms of spherical Bessel functions: $y_1(kr) = [G(kr) \exp(\rmi kr) + G^*(kr) \exp(-\rmi kr)]/2=-\cos (kr)/(kr)^2 - \sin (kr)/(kr)$ and $j_1(kr) = [-G(kr)\exp(\rmi kr) +G^*(kr) \exp(-\rmi kr)]/(2i) = \sin(kr)/(kr)^2 - \cos(kr)/(kr)$.}, and $G(x)=\rmi/x-1/x^2$.
Here $C_e^T$, $C_r^T$, $C_e^L$ and $C_r^L$ are dimensionless complex valued constants. The term with $C_e^T$ will give an expanding T-wave, the term with $C_r^T$ a reflected (spherical incoming) T-wave. Similarly the terms $C_e^L$ and $C_t^L$ give expanding and reflected L-waves in the shell. For the (infinite) external domain there is only a `transmitted expanding' (subscript `t') wave with 
\begin{equation}
\begin{aligned} \label{eq:u_out}
    \bs{u}^{\rmout} = \left[ -r \frac{\rmd}{\rmd r} \left(\frac{h^{\rmout}}{r}\right) - 2 \frac{h^{\rmout}}{r} + \frac{\phi^{\rmout}}{r}\right] \bs{u}^0 + \frac{\rmd}{\rmd r}\left(\frac{h^{\rmout}}{r} +\frac{\phi^{\rmout}}{r} \right) \frac{\bs{x} \cdot \bs{u}^0}{r} \bs{x},
    \end{aligned}
\end{equation}
\begin{equation}
\begin{aligned}
h^{\rmout}(r)=h_t(r)=-a C_t^T \exp(\rmi k_T^{\rmout}r) G(k_T^{\rmout} r), \\  \phi^{\rmout}(r)=\phi_t(r)=-a C_t^L \exp(\rmi k_L^{\rmout}r) G(k_L^{\rmout} r) 
\end{aligned}
\end{equation}
with $k_T^{\rmout}$ and $k_L^{\rmout}$ the parameters for the external domain and $C_t^T$ and $C_t^L$ dimensionless constants. 

There are six unknown parameters: $C_e^T$, $C_r^T$, $C_e^L$, $C_r^L$, $C_t^T$ and $C_t^L$ which can be determined by matching the $u_i$ components of the displacement at $r=a$ (two equations), the displacement at $r=b$ (two equations), and finally the continuity of the shear stress at $r=b$ (two equations). The details of how to get these six parameters are described in \ref{App:AppA}.

An alternative way of getting the solution using spherical coordinates and spherical Bessel and Hankel functions of the first kind is given in \ref{App:AppB}. Both approaches are equivalent and give the same result. This way of solving the problem corresponds more to the classical mathematical approach, yet the constants of \ref{App:AppA} correspond directly to the `emitted' and `reflected' waves in the elastic layer, while this is not explicitly the case in the approach of \ref{App:AppB}.

\begin{figure*}[t]
\begin{center} 
\includegraphics[width=0.32\textwidth]{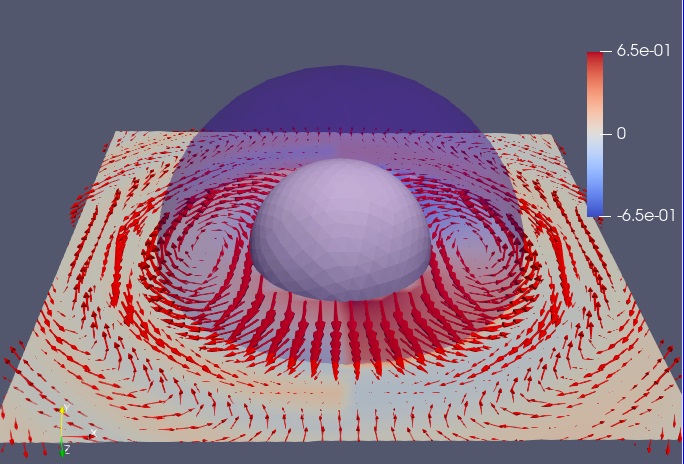}
\includegraphics[width=0.32\textwidth]{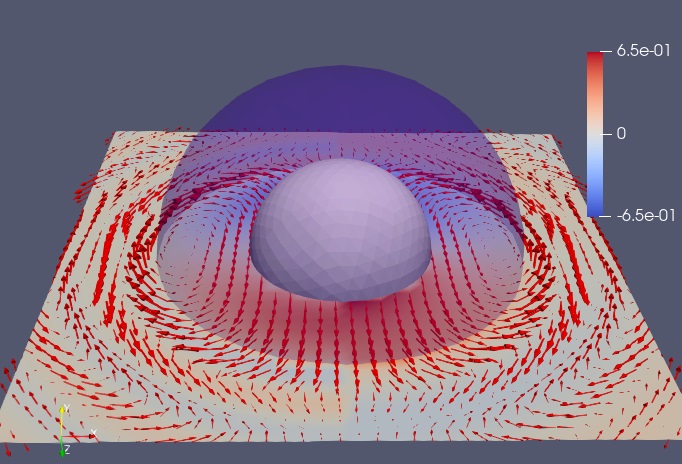}
\includegraphics[width=0.32\textwidth]{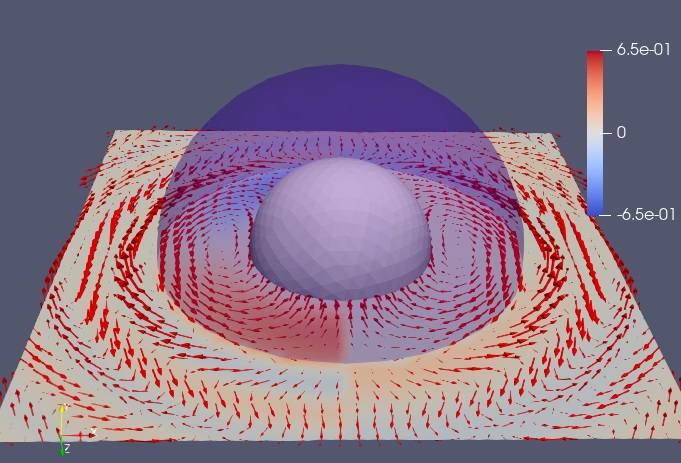}\\
\includegraphics[width=0.32\textwidth]{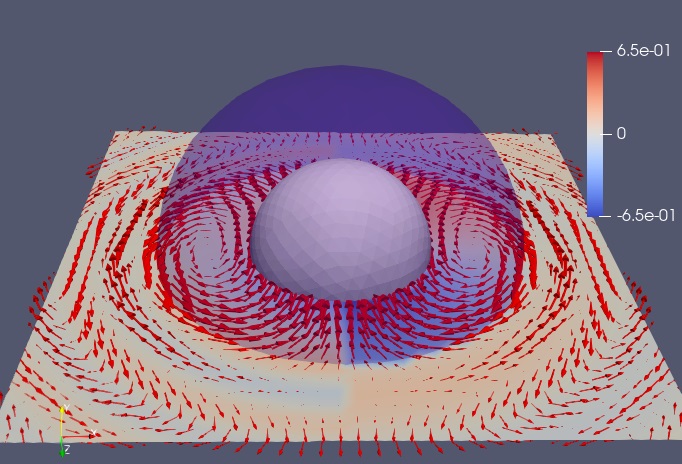}
\includegraphics[width=0.32\textwidth]{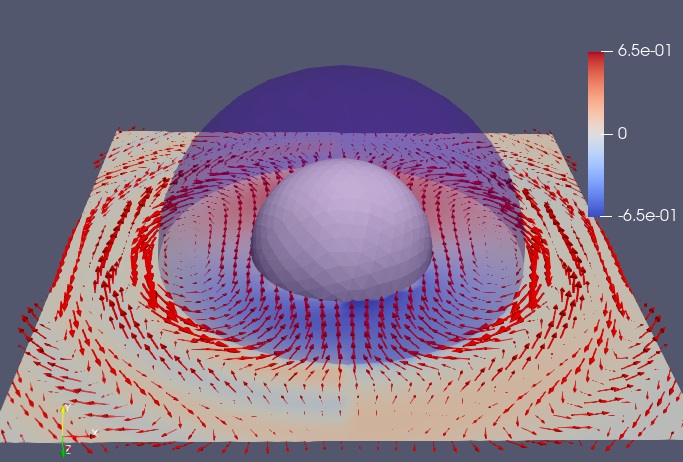}
\includegraphics[width=0.32\textwidth]{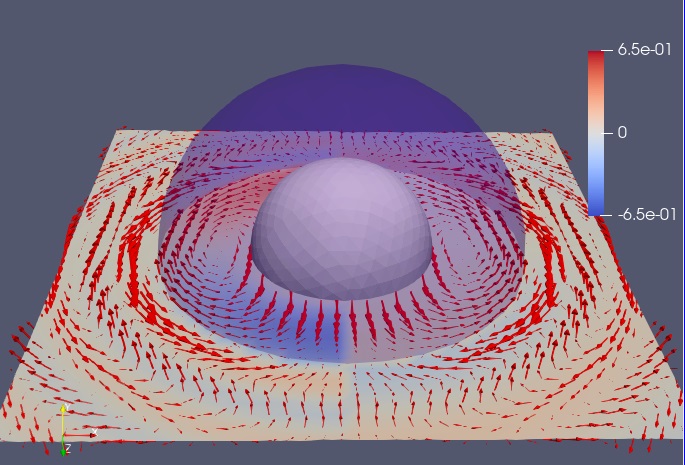}
\caption{Inner rigid core sphere `vibrating' periodically with amplitude $U^0$ in another concentric spherical shell ``sh'', the total embedded in an infinite external outer domain ``out'' with $b/a=2.0$. Parameters: $k_T^{\rmsh} a=2.5$, $k_L^{\rmsh} a=1.0$, $k_T^{\rmout}a=8.0$, $k_L^{\rmout}a=3.0$, $\rho_{\rmout}/\rho_{\rmsh}=1.0$. The sphere oscillates from back to front of the figure. On the horizontal plane the total displacement vectors are plotted. A complicated pattern is formed due to the interaction of L and T waves. On the left of the plane the function $h(r) \cos(\theta)$ is plotted while on the right hand side $\phi(r) \cos(\theta)$ is plotted in color. Here time-snapshots are shown at 0/6, 1/6, 2/6, 3/6, 4/6 and 5/6 times the oscillation cycle. A movie file is available for this case showing 30 time frames. }\label{Fig:sphereShell1}
\end{center}
\end{figure*}

\section{Results} \label{sec:results}

Some screenshots of the solution with parameters $\rho_{\rmout}/\rho_{\rmsh} =1.0$, $b/a=2.0$, $\boldsymbol u^0 = (0,0,U^0)$, $k_T^{\rmsh} a =2.5$, $k_L^{\rmsh} a=1.0$, $k_T^{\rmout}a=8.0$ and $k_L^{\rmout} a=3.0$ are shown in Fig.~\ref{Fig:sphereShell1}. Since any solution $\bs{u} \exp(i \alpha)$, with $\alpha$ a phase factor, is also a solution of the problem, we can easily reconstruct the solution in the time domain, by choosing appropriate values for $\alpha$ for each time step. The inner core sphere vibrates front to back.  The shell/outer medium boundary is indicated in transparent blue. A $40 \times 40$ grid is chosen on the horizontal plane and the displacements are indicated on this grid with arrows. An intricate pattern of displacements can be observed. Since $k_T^{\rmout} > k_T^{\rmsh}$ and $k_L^{\rmout} > k_L^{\rmsh}$, the waves in the outer domain are more densely packed than in the shell. In the outer domain waves can be seen to travel towards infinity, while in the shell complex interference patterns appear due to the interaction of the emitted and reflected waves. 

\begin{figure*}[t]
\begin{center}  
\includegraphics[width=0.95\textwidth]{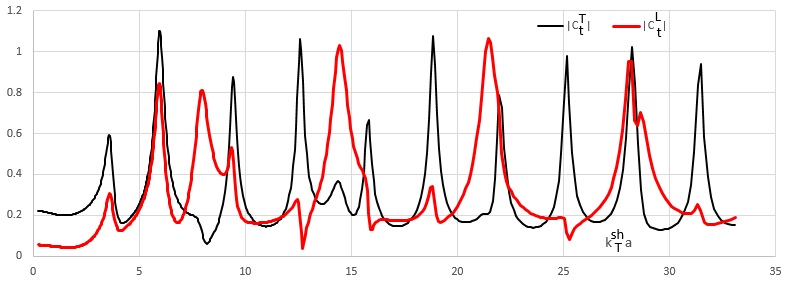}
\caption{Frequency response curve for a sphere with a shell. Transmitted T and L-coefficients $|C^T_t|$ and $|C^L_t|$ (i.e. into the external domain) as a function of the parameter $k_T^{\rmsh} a$. Here, we keep the material parameters the same, but the oscillation frequency $\omega$ is changed. The parameter are $\rho_{\rmout}/\rho_{\rmsh} =3.0$, $b/a=2.0$, $\boldsymbol u^0/U^0 = (0,0,1)$. Take initially $k_T^{\rmsh} a =4.5$, $k_L^{\rmsh} a=2.0$, $k_T^{\rmout}a=2.0$ and $k_L^{\rmout} a=1.0$. Then change the frequency, thus multiply each $k$ by the same value and recalculate all constants. Note the rather chaotic character of this graph, with many maximum and minimum values for both the T and L-coefficients, however, not at the same frequencies. The `spectrum' shows remarkable peaks and valleys especially compared to the case when no shell as presented in Fig.~\ref{Fig:sol_noShell}. } \label{Fig:solC_t}
\end{center}
\end{figure*}

\begin{figure*}[t]
\begin{center} 
\includegraphics[width=0.95\textwidth]{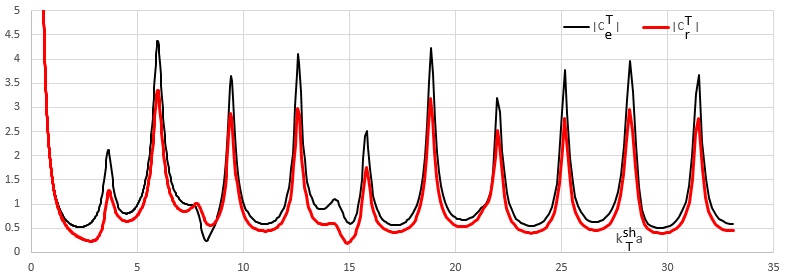}
\caption{Frequency response for a sphere with shell. As Fig.\ref{Fig:solC_t}, but now for the emitted and reflected T-coefficients $|C^T_e|$ and $|C^T_r|$ (in the shell) as a function of $k_T^{\rmsh} a$.  The peaks and valleys are mostly overlapping, but not always. Note that both coefficients diverge at $k_T^{\rmsh} a=0$.}\label{Fig:sol_erT}
\end{center}
\end{figure*}

\begin{figure*}[t]
\begin{center} 
\includegraphics[width=0.95\textwidth]{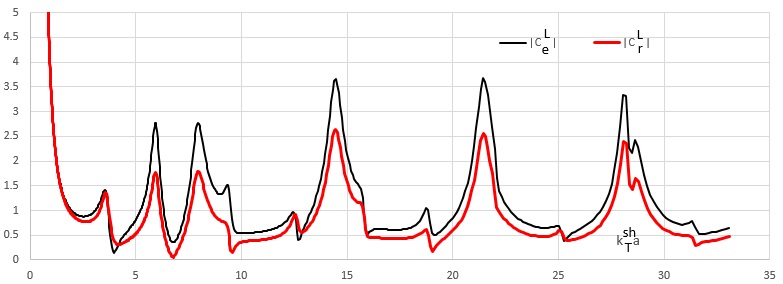}
\caption{Frequency response for a sphere with shell. As Fig.\ref{Fig:solC_t}, but now for the emitted and reflected L-coefficients: $|C^L_e|$ and $|C^L_r|$ (in the shell) as a function of $k_T^{\rmsh} a$. Note that both coefficients diverge at $k_T^{\rmsh} a=0$. } \label{Fig:sol_erL}
\end{center}
\end{figure*}

Next we wonder what will happen if we keep the physical system the same, but change the oscillation frequency $\omega$. Since $k=\omega/c$, this is essentially the same as multiplying all $k$ values (both L and T and the shell and the outer domain) by the same value. We choose an example with the following  parameters: $\rho_{\rmout}/\rho_{\rmsh} =3.0$, $b/a=2.0$, $\boldsymbol u^0 = (0,0,1)$. Take initially $k_T^{\rmsh} a =4.5$, $k_L^{\rmsh} a=2.0$, $k_T^{\rmout}a=2.0$ and $k_L^{\rmout} a=1.0$. Then gradually increase (or reduce) the frequency, i.e. multiply (or divide) each $k$ by $1.005$ and recalculate all $C$'s until $k_T^{\rmsh} a=33$. The results for the transmitted coefficients (in the outer domain) $|C_t^T|$ and $|C_t^L|$ are shown in Fig~\ref{Fig:solC_t}. A complicated spectrum of peaks and valleys appears. For some values of $k_T^{\rmsh} a$,  $|C_t^T|$ is near zero, while for other values $|C_t^L|$ becomes near zero. The emitted and reflected coefficients $|C_e^T|$ and $|C_r^T|$ for the transverse waves in the shell are shown in Fig.~\ref{Fig:sol_erT}. The first peaks appear near $k_T^{\rmsh}=3.6$. Note that both $|C_e^T|$ and $|C_r^T|$ tend towards infinity when $k_T^{\rmsh} a=0$. Finally, the emitted and reflected coefficients $|C_e^L|$ and $|C_r^L|$ for the longitudinal waves in the shell are shown in Fig.~\ref{Fig:sol_erL}. Again both $|C_e^L|$ and $|C_r^L|$ tend towards infinity when $k_T^{\rmsh} a=0$. 

\begin{figure*}[!ht]
\begin{center}  
\subfloat[]{\includegraphics[width=0.4\textwidth]{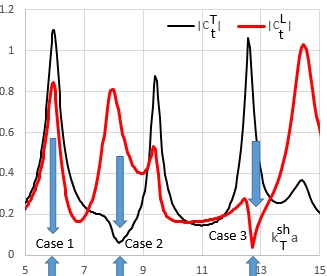}} \quad\quad\quad
\subfloat[]{\includegraphics[width=0.47\textwidth]{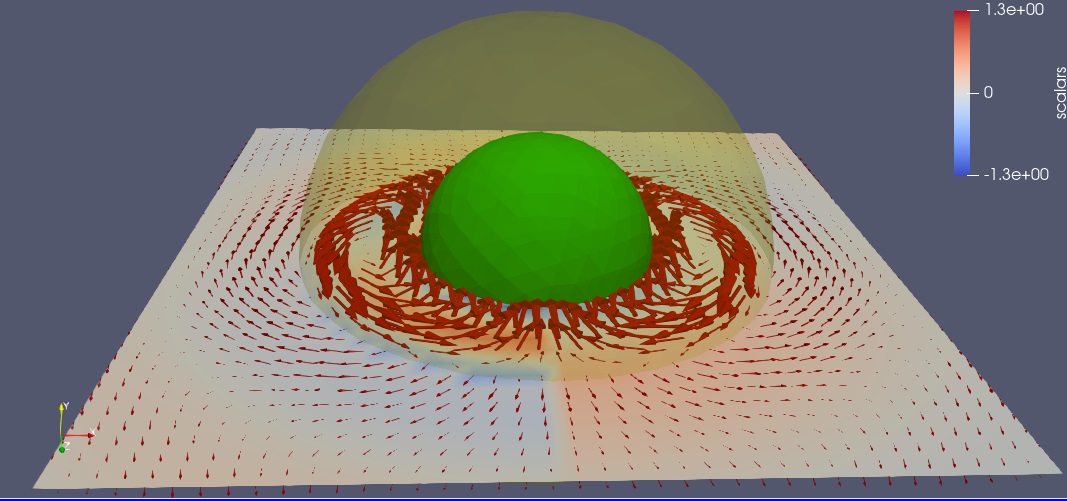}}\\
\subfloat[]{\includegraphics[width=0.47\textwidth]{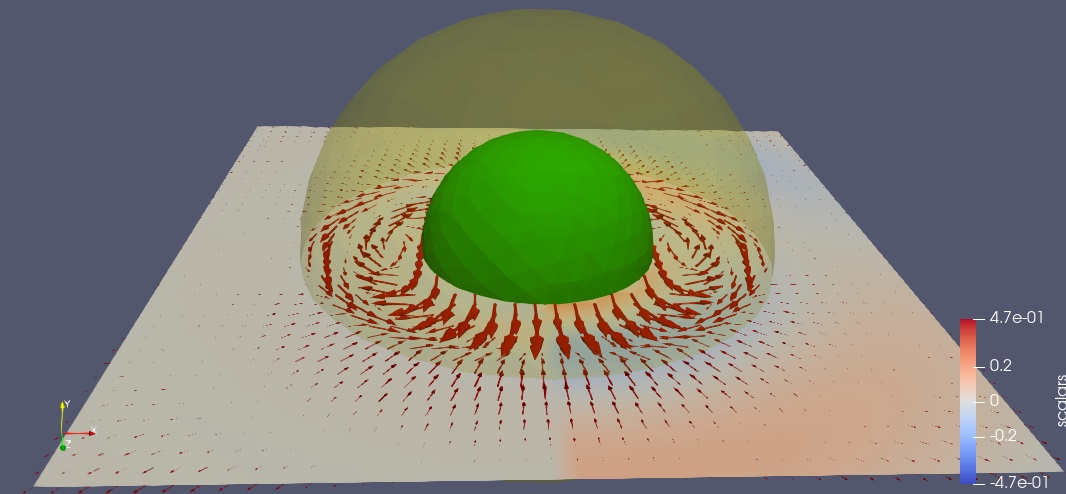}}\quad
\subfloat[]{\includegraphics[width=0.47\textwidth]{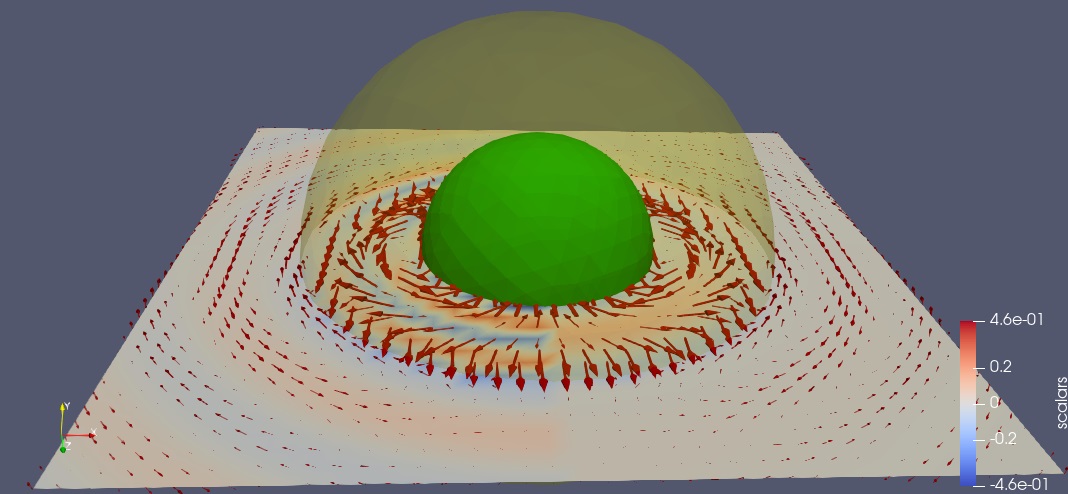}}
\caption{(a) Zoom in of Fig.~\ref{Fig:solC_t} with the three selected cases indicated by arrows. Vector plots in the horizontal plane: (b) Case 1: Both T-waves and L-waves are generated in the external domain. (c) Case 2: Mainly L-waves occur in the external domain. (d) Case 3: Mainly T-waves appear in the external domain. The function $h(r)\cos(\theta)$ is plotted on the left hand side of the horizontal plane and $\phi(r) \cos(\theta)$ is shown on the right hand side. It is thus possible, by a clever combination of materials and frequency to generate mainly L or mainly T waves or a combination of both. 
} \label{Fig:Cases}
\end{center}
\end{figure*}

Let us investigate what exactly happens in these peaks and valleys by investigating three cases. Based on Fig.~\ref{Fig:solC_t} or the zoom-in shown in Fig.~\ref{Fig:Cases}(a), around $k_T^{\rmsh} a =5.98$, both $|C^T_t|$ and $|C^L_t|$ are near a peak value. We will call this Case 1. Thus for Case 1, $k_L^{\rmsh} a = 5.98\times2.0/4.5$, $k_T^{\rmout} a =5.98\times2.0/4.5$ and $k_L^{\rmout} a =5.98/4.5$ (scale all wavenumbers with the same amount). The cases are indicated with blue large arrows for clarity in Fig.~\ref{Fig:Cases}(a). In Fig.~\ref{Fig:Cases}(b), the displacement pattern is shown with red arrows. We can see that both T- and L-waves appear in the outer domain. Around $k_T^{\rmsh} a = 8.18$ the constant $|C^T_t|$ becomes very small while $|C^L_t|$ becomes large. Take $k_L^{\rmsh} a = 8.18\times2.0/4.5$ and $k_T^{\rmout}a=8.18\times2.0/4.5$ and $k_L^{\rmout}a=8.18/4.5$, we call this Case 2. In Fig.~\ref{Fig:Cases}(c), the displacement pattern is shown with red arrows, we see that they are all pointing radially in or outwards, indicating that mainly L-waves occur in the outer domain. Finally, for Case 3, we take the value $k_T^{\rmsh} a = 12.70$, where a minimum in $|C^L_t|$ occurs, again we multiply all wavenumbers by the same amount as in Cases 2 and 3. Now we clearly see a T-wave in the outer domain (all displacement vectors in Fig.~\ref{Fig:Cases}(d) are 90 degrees rotated when compared to Case 2). In all three cases, $\phi(r)\cos(\theta)$ is shown on the right hand side of the horizontal plane and $h(r) \cos(\theta)$ is shown on the left hand side in color. 

The constants $C_e^T$ and $C_r^T$, representing the emitted and reflected transverse waves in the shell, are shown in Fig.~\ref{Fig:sol_erT} and the $C_e^L$ and $C_r^L$ constants in Fig.~\ref{Fig:sol_erL} which show the emitted and reflected longitudinal waves in the shell. 

\section{Discussion} \label{sec:discussion}

Note that $\bs{u^0}$ can be different from the (real valued) $\bs{u^0}/U^0=(0,0,1)$, it could assume a complex value as well (as long as it is a constant). For example $\bs{u^0}/U^0=(\rmi,0,1)/\sqrt 2$ will give a circularly vibrating sphere (not shown here).

The following six non-dimensional parameter space can be distinguished for the shell case: $b/a$, $k_T^{\rmsh} a$, $k_T^{\rmsh}/k_L^{\rmsh}$, $k_T^{\rmsh}/k_T^{\rmout}$, $k_T^{\rmsh}/k_L^{\rmout}$ and $\rho_{\rmout}/\rho_{\rmsh}$ (or any combination of these parameters). 
For a typical $a=1$ mm application, with $c_L = 6000$ m/s and $\rho=1000$ kg/m$^3$, a value of $k_L^{\rmsh} a= 1.0$ would correspond to a frequency of $2 \pi \omega = 1$ MHz and for $k_L^{\rmsh} a =100$, one would need 100 MHz, a frequency that is now becoming available in acoustic transducers~\citep{Fei2016}. For an object with typical size $a=1$ m, the frequency will be 1 kHz for $k_L^{\rmsh} a=1$ and 100 kHz for $k_L^{\rmsh} a=100$. 

The current framework can easily be extended to multiple shells. For a core with a single shell, we had to solve a $6 \times 6$ matrix, with every additional shell we will have to add 4 more equations, thus a $10 \times 10$ matrix for a two-shells system for example. It is also possible to calculate the stresses caused by the movement of the sphere, although we have not shown them here. Although outside the scope of the current analytical solution, a real system could easily be built for example by embedding a steel sphere in an elastic material and exciting it by magnetic means, for example possibly to convert electrical to mechanical energy and to generate heat remotely via magnetic stimuli. On the other hand, the study of this relatively simple system, yet with complex behavior, opens the way to further study and better understand real systems with their associated resonances, noise generation, fatigue and failure behavior, frequency responses etc.

As we have shown the current analytical solution exhibits non-trivial behavior. It has rich physical detail, for example the presence of both longitudinal and transverse waves, including interference between outgoing and reflected waves and is therefore ideally suited to test numerical solutions, for example those generated by finite element or boundary element codes. In \ref{App:BEM}, we have used the analytical solution to test a boundary element code based on the framework developed by~\cite{Rizzo1985}. Excellent agreement is achieved when the numerical solution is compared to the theory. 

\subsection{Vibrating sphere without a shell}

A solution for a vibrating sphere without a shell was previously given by~\cite{KlaseboerJElast2019}\footnote{
In order to get back the same solutions the constants used there should be replaced by
with $C^T_e = - 2 c_1$ and $C^L_e = k_L^2 a^2 [2 c_1/(k_T^2 a^2) + c_2]$, where $c_1$ and $c_2$ are the constants used by \cite{KlaseboerJElast2019}.}, which is a special case of the current work when the shell material is set to be the same as the outer material. For such a case which is the equivalent to no shell at all, the number of parameters mentioned in the previous section (six) reduces to two, namely $k_T^{\rmout} a$ and $k_T^{\rmout}/k_L^{\rmout}$. 

\begin{figure*}[t]
\begin{center} 
\includegraphics[width=0.95\textwidth]{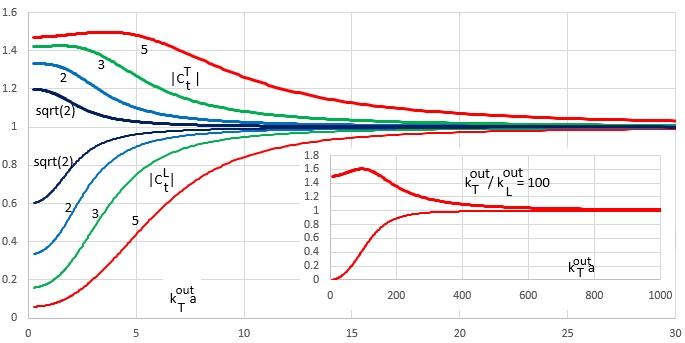}
\caption{Frequency response curves for a sphere with no shell. $|C^T_t|$ (upper curves) and $|C^L_t|$ (lower curves) for various $k_T^{\rmout}/k_L^{\rmout}$ ratios, from $k_T^{\rmout}/k_L^{\rmout}=\sqrt{2}$ (the smallest this ratio can be) to $k_T^{\rmout}/k_L^{\rmout}=2, 3, 5$ and $k_T^{\rmout}/k_L^{\rmout}=100$ in the inset (going towards the in-compressible limit). Note the smoothness of the curves when the parameter $k_T^{\rmout} a$ is changed (which is essentially the same as changing the driving frequency), which is in stark contrast to the curves shown in Fig.~\ref{Fig:solC_t}.} \label{Fig:sol_noShell}
\end{center}
\end{figure*}

When we keep the material parameters constant and change the vibration frequency (thus changing $k_T^{\rmout} a$ and keeping $k_T^{\rmout}/k_L^{\rmout}$ constant), we can calculate $|C_t^T|$ and $|C_t^L|$. The results are plotted in Fig.~\ref{Fig:sol_noShell}. When compared to Fig.~\ref{Fig:solC_t}, the smoothness of the curves in Fig.~\ref{Fig:sol_noShell} is immediately noticed. The larger the $k_T^{\rm\rmout}/k_L^{\rmout}$ ratio is, the smaller the longitudinal parameter $|C_t^L|$ becomes for low $k_Ta$ values $(k_Ta <<1)$. But $|C_t^T|$ and $|C_t^L|$ both converge towards a value of 1.0 for larger $k_T^{\rmout} a$ values.

For a vibrating sphere with no shell, it seems not possible to have a zero L or zero T contribution according to Fig.~\ref{Fig:sol_noShell}. The only possibility to generate a near zero L-wave is to reduce the frequency to near zero values. For larger frequencies (thus larger $k_T^{\rmout} a$ values), all curves tend towards $|C_L^t|=|C_T^{t}|=1$. The $|C_t^L|$ curves all seem to be monotonously increasing. For $k_T^{\rmout}/k_L^{\rmout}=\sqrt{2}$ and $k_T^{\rmout}/k_L^{\rmout}=2$, the $|C_t^T|$ curves monotonously decrease. However, for $k_T^{\rmout}/k_L^{\rmout}=3$ and onward, these curves show a maximum value of $|C_t^L|$. For $k_T^{\rmout}/k_L^{\rmout}=3$, the maximum $|C_t^L|$ is at $k_T^{\rmout}a = 1.492$ with a value of $1.426$ ($|C_t^L|$ is $1.421$ at $k_T^{\rmout}a$ near zero). The maximum $|C_t^L|$ appears later for $k_T^{\rmout}/k_L^{\rmout}=5$ at $k_T^{\rmout}a = 3.592$ with a value of $1.498$. The maximum $|C_t^L|$ shifts to larger and larger values of $k_T^{\rmout}a$ when the ratio $k_T^{\rmout}/k_L^{\rmout}$ increases further, but does not seem to go significantly above a value of $1.6$. For example, in the inset the curves for $k_T^{\rmout}/k_L^{\rmout}=100$ are shown where the maximum $|C_t^L|$ occurs around $k_T^{\rmout}a = 95.3$ with $1.607$. For even larger $k_T^{\rmout}/k_L^{\rmout}=1000$ (not shown) the maximum $|C_t^L|=1.612$ and occurs at $k_T^{\rmout}a = 970$ approximately\footnote{Fig.~\ref{Fig:sol_noShell} was generated by setting $b/a=2$, $k_L^{\rmsh}a=k_L^{\rmout} a$, $k_T^{\rmsh} a = k_T^{\rmout} a$ and $\rho_{\rmsh}=\rho_{\rmout}$ (thus setting the material of the shell identical to that of the outer domain). We then get $C_r^L=0$, $C_r^T=0$, $C_e^L=C_t^L$ and $C_e^T=C_t^T$, as it should be.}.

\begin{figure*}[t]  
\begin{center} 
\includegraphics[width=0.75\textwidth]{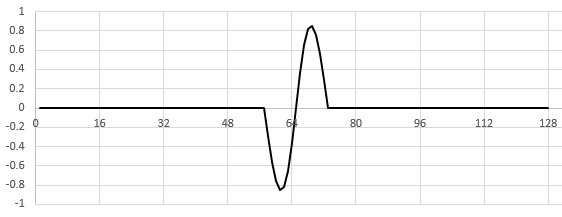}
\caption{The used -/+ pulse for the FFT transform with width $W=5a$ and 128 points. The values from $i=57$ to $72$ are non-zero with a minus/plus pulse centered at $i=65$ as $\text{DATA}[i]=2(i+N-N_w/2)-1]=\sin(2x) \exp(-\alpha x/2)$ with $x = (i-1-N_w/2)\pi/N_w$, with $N=64$, $N_w=N/4$ and $\alpha=0.1$. Since the wave is antisymmetric, the lowest frequency of the 65 frequencies in the FFT corresponds to $k_T^{\rmsh}a=1.256$ and the highest to $k_T^{\rmsh}a=80.425$ (there is no need to calculate $k_T^{\rmsh}a=0$). 
}
 \label{Fig:FFT1Pulse}
\end{center}
\end{figure*}

\subsection{Pulsed time domain solutions using the Fast Fourier Transform}

Now that the response for each frequency can be calculated we can use the Fast Fourier Transform (FFT)~\citep{Bedford1994} to get the solution for the displacements, at each location and for each time instant, if we assume the core is exhibiting a pulsed vibration. The minus/plus pulse used is shown in Fig.~\ref{Fig:FFT1Pulse}. We have deliberately chosen an antisymmetric pulse in order to avoid issues with non vanishing displacements associated with $k_T^{\rmsh} a = 0$. The standard FFT procedure given in the book Numerical Recipes was used~\citep{Numrep}. 

\begin{figure*}[t]  
\begin{center} 
\includegraphics[width=0.32\textwidth]{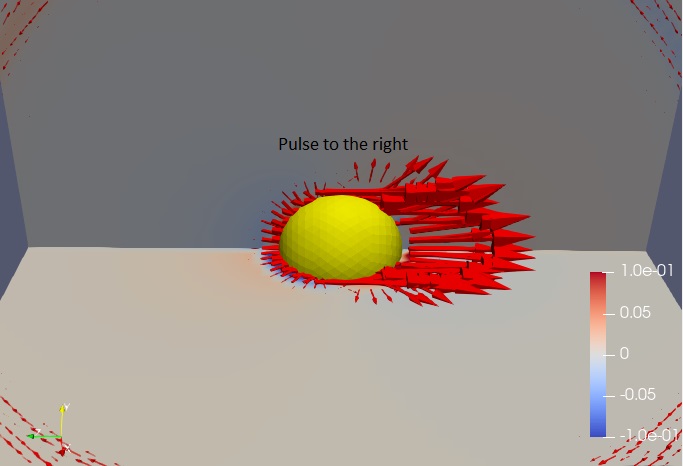}
\includegraphics[width=0.32\textwidth]{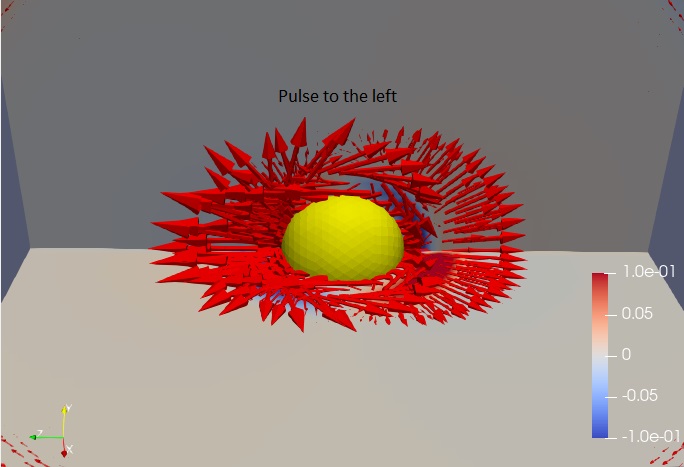}
\includegraphics[width=0.32\textwidth]{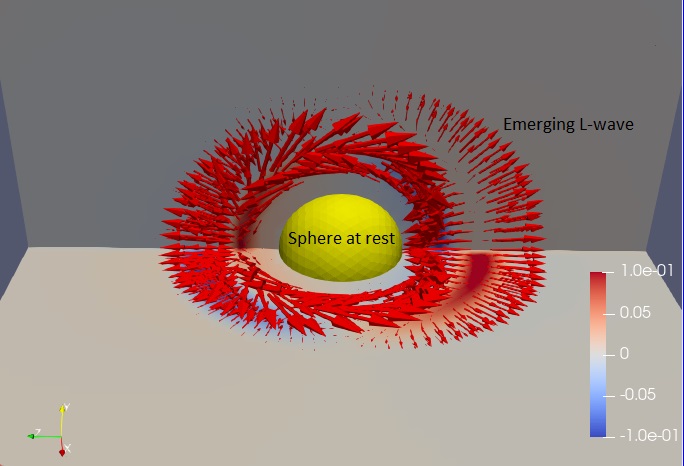}\\
\includegraphics[width=0.32\textwidth]{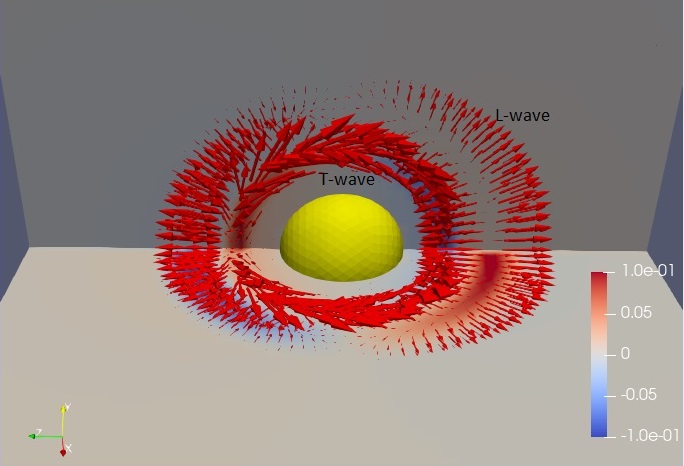}
\includegraphics[width=0.32\textwidth]{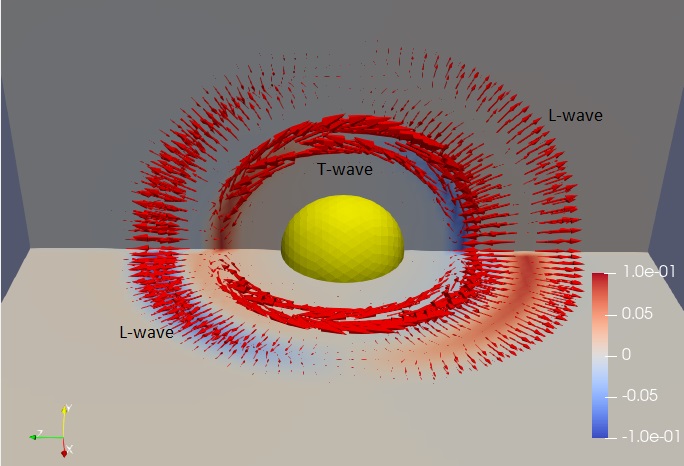}
\includegraphics[width=0.32\textwidth]{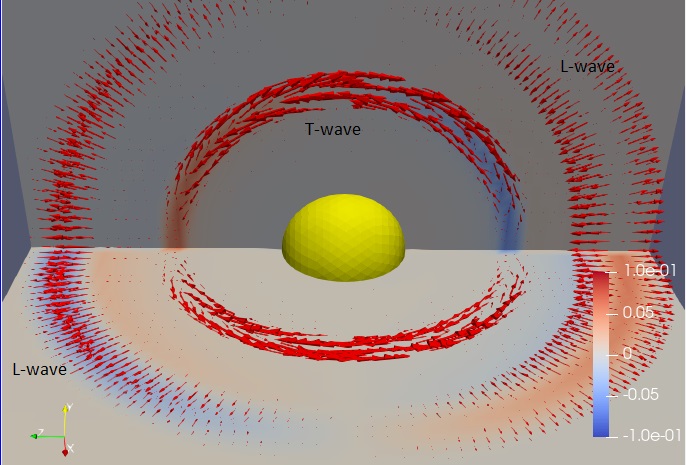}
\caption{Screenshots at different times, obtained with a FFT-transform of a single -/+ pulse (as shown in Fig.~\ref{Fig:FFT1Pulse}) for the case of a sphere with no shell. The sphere first moves to the right and then to the left before it stops moving. The resulting displacement patterns (in vectors) show the separation of the L-waves and the T-waves in the 4th and 5th image onward. The L-waves travel twice as fast for this particular case ($k_T^{\rmout}/k_L^{\rmout}=2.0$). On the horizontal plane the scalar function $\phi(r) \cos \theta$ is also given, while on the vertical plane $h(r) \cos \theta$ is plotted in color. A movie file is available for this case. } \label{Fig:FFT1}
\end{center}
\end{figure*}

In Fig.~\ref{Fig:FFT1} a typical example of the separation of the T and L waves (where the L waves travel at twice the speed as the T waves) for the case with no shell associated with the pulse given in Fig.~\ref{Fig:FFT1Pulse}. Initially the T and L waves are interfering with each other, until from Frame 4 onward, the L wave (most outer wave) clearly separates from the T wave (inner wave). The material parameters for this case are $k_T^{\rmout}/k_L^{\rmout}=2.0$.

\begin{figure*}[!ht]  
\begin{center} 
\includegraphics[width=0.32\textwidth]{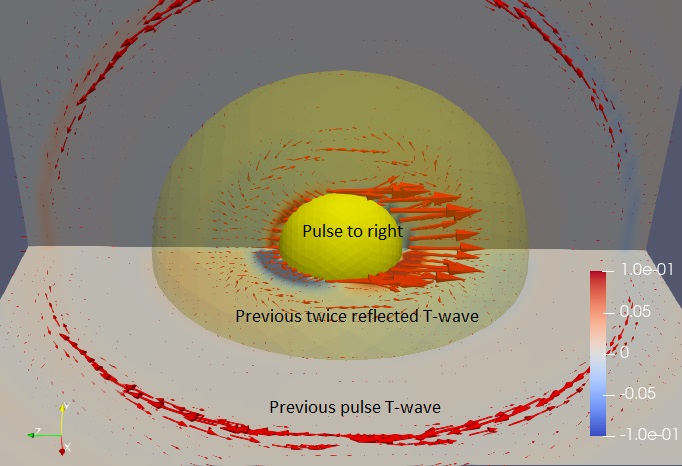}
\includegraphics[width=0.32\textwidth]{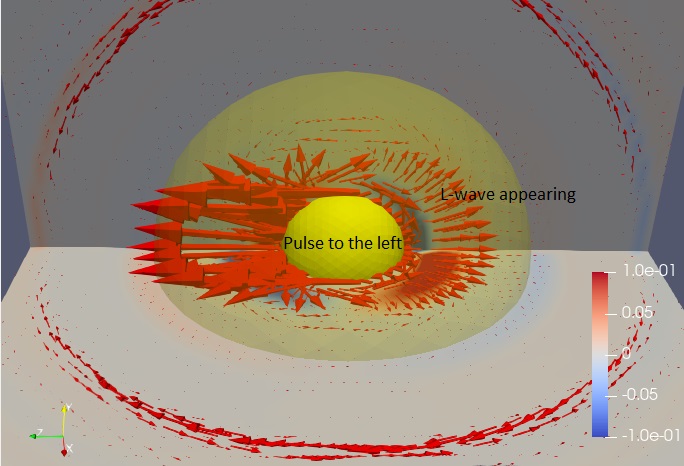}
\includegraphics[width=0.32\textwidth]{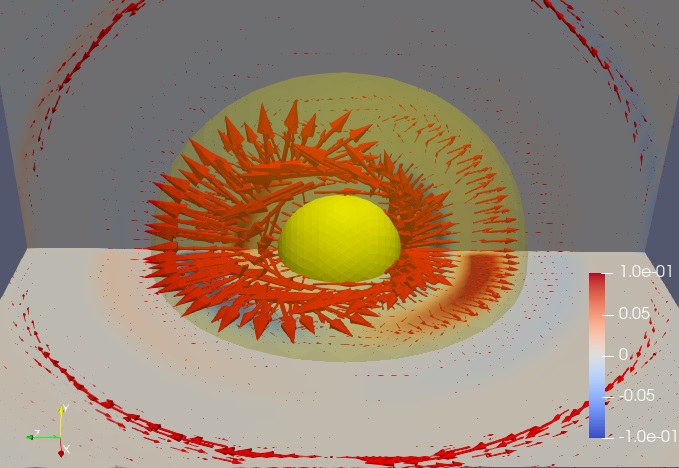}\\
\includegraphics[width=0.32\textwidth]{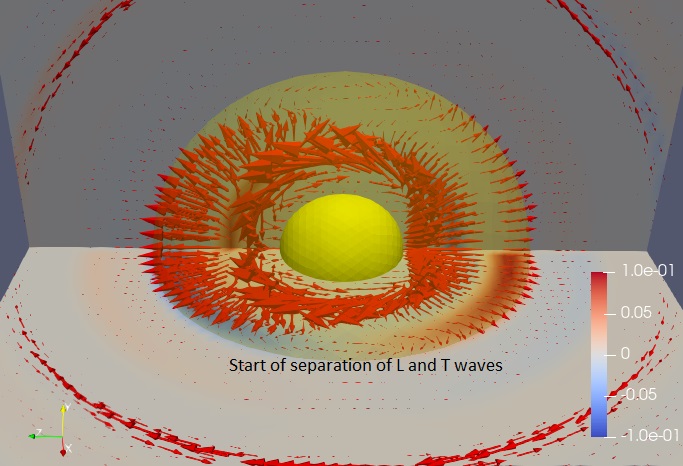}
\includegraphics[width=0.32\textwidth]{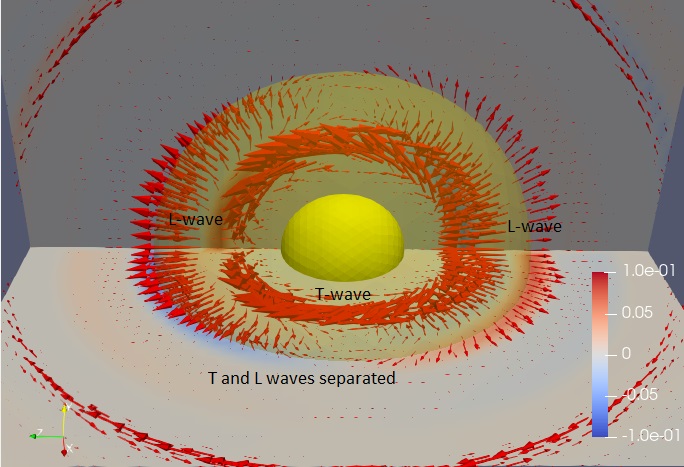}
\includegraphics[width=0.32\textwidth]{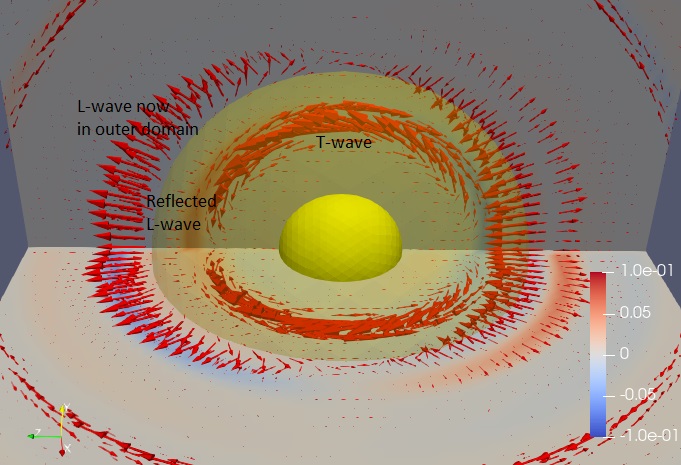}\\
\includegraphics[width=0.32\textwidth]{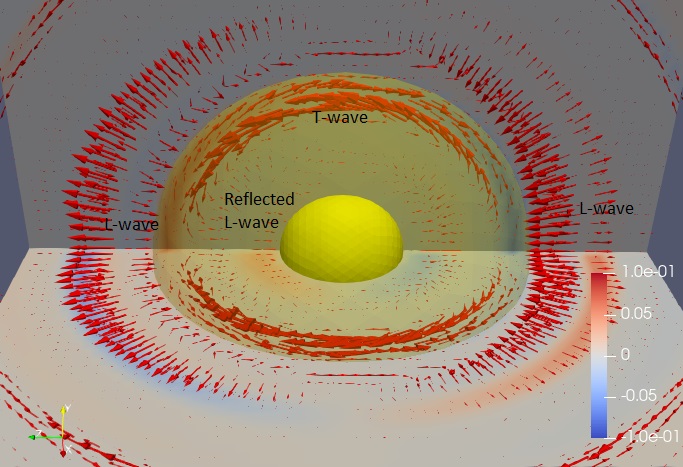}
\includegraphics[width=0.32\textwidth]{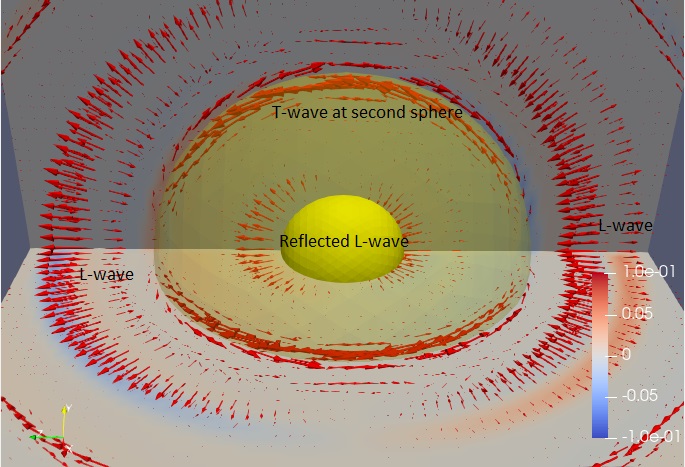}
\includegraphics[width=0.32\textwidth]{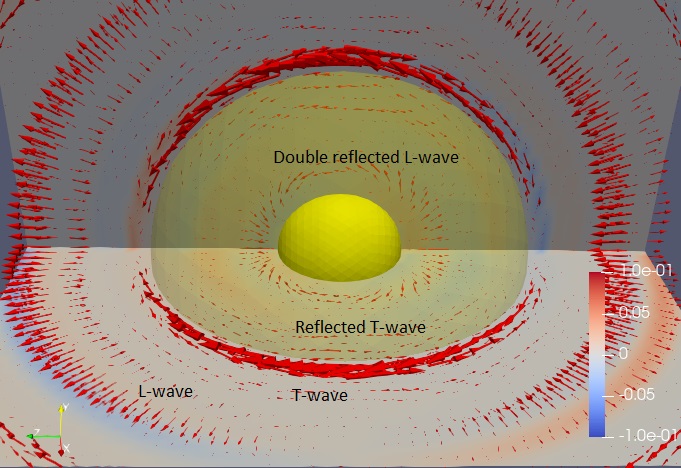}\\
\includegraphics[width=0.32\textwidth]{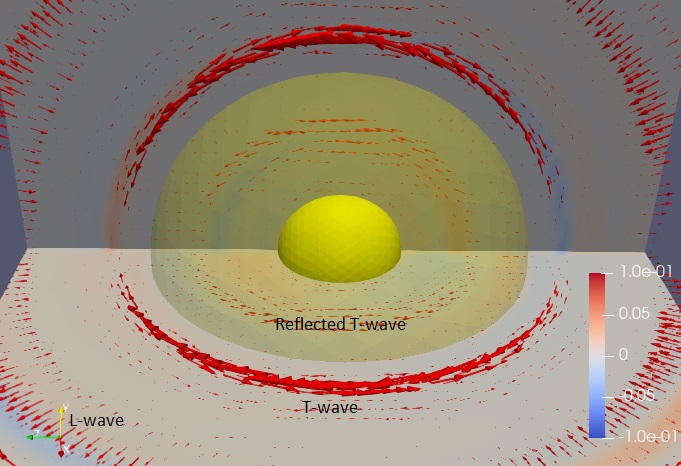}
\includegraphics[width=0.32\textwidth]{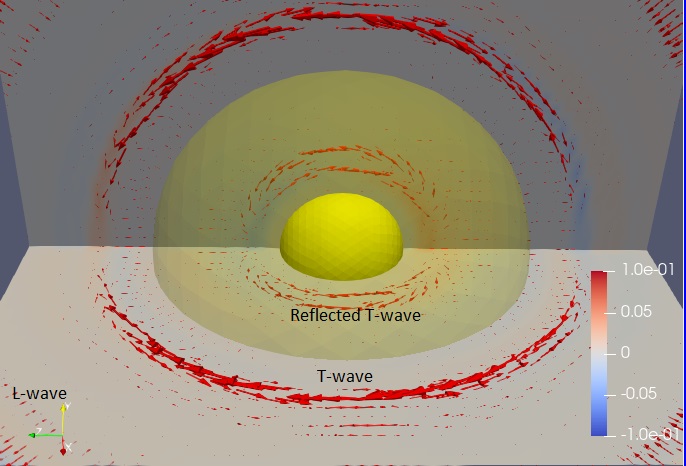}
\includegraphics[width=0.32\textwidth]{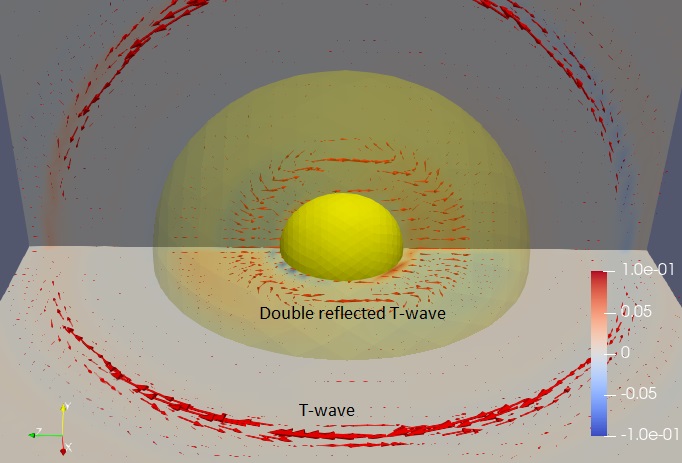}
\caption{Screenshots at different times, obtained with a FFT-transform of a single -/+ pulse (as shown in Fig.~\ref{Fig:FFT1Pulse}) for the case with a shell with $b/a=3$ (the second sphere is indicated in transparent yellow color). Multiple reflections and double reflections of the T and L-waves can be observed, which obviously do not occur for the case with no shell of Fig.~\ref{Fig:FFT1}. A movie file is available for this case.} \label{Fig:FFTShell}
\end{center}
\end{figure*}

Another case with a shell is shown next. The parameter chosen are $b/a=3$, $k_T^{\rmsh}a = 4.0$, $k_T^{\rmsh}/k_L^{\rmsh}=2.0$, $k_T^{\rmsh}/k_T^{\rmout}=4/7$, $k_T^{\rmsh}/k_L^{\rmout}=4/3$ and $\rho_{\rmout}/\rho_{\rmin}=1.0$. This parameter set will give results with not too many reflections. The results are shown in Fig.~\ref{Fig:FFTShell}. The shell/outer boundary is indicated with a transparent yellow sphere. Expanding waves can be observed caused by the pulse which is the same as the one used in Fig.~\ref{Fig:FFT1}. Also, reflected waves from the shell/outer boundary can be clearly distinguished and even doubly reflected waves (i.e. reflected waves that reflect once more on the inner rigid core sphere). In order to more easily differentiate the expanding and reflecting waves, they are indicated in each frame.

\section{Conclusions} \label{sec:conclusion}

The analytical solution for the dynamic elastic problem of a vibrating rigid sphere surrounded by an elastic shell, the total being immersed in another infinite medium is presented. The solution shows some surprisingly unexpected physics with various peaks for both the longitudinal (L) and transverse (T) response when the frequency of the vibration is changed. These do not appear for the simpler case of a sphere without an elastic shell layer, where the frequency response is a smooth line. In practice, this means that almost pure L or T waves can be generated by carefully choosing the material parameters and the frequency of the vibration of the core sphere. This complex behavior, which is not present for spheres without a shell, appears very similar to the unique properties that can be observed with mechanical metamaterials, see \cite{Kelkar2020} or \cite{Wang2014}. 

Since all the responses for multiple frequencies can be easily obtained in the frequency domain, we can use the FFT framework to predict the response to a pulsed vibration in the time domain. Some examples are shown for a narrow pulse, which shows the separation of the L and T waves which move out radially as clearly distinctive pulses after some time has passed.

In this article we have just scratched the surface regarding the multitude of possible solutions; there are six dimensionless parameters that can be varied in the core-shell vibration problem. The solution could be considered as the first approximation of an oscillating body in an elastic material. 

The analytical solutions can also be used as benchmark cases to test numerical solutions obtained with for example the finite element method (where boundary conditions at infinity are not easy to implement) or a boundary element method (which have hyper-singular integrals that need to be treated with extreme precaution), even in the time domain when the fast Fourier transform framework is used (such as in the example of Fig.~\ref{Fig:FFTShell}). 

The implementation of the solution is relatively straightforward, without any infinite sums or other mathematical difficulties. The codes (in Fortran language) used to generate the plots in this article are available from the authors on request. 

\section*{Acknowledgments}
Q.S. was supported by the Australian Research Council (ARC) through Grants DE150100169, FT160100357 and CE140100003.

\appendix
\section{The solution in terms of vector - tensor notation}\label{App:AppA}
The conditions on the displacement at $r=a$ and $r=b$ give (by equating the terms with $\bs{x}$ and $\bs{u}^0$ in $\bs{u}^{\rmsh}=\bs{u}^0$ at $r=a$ and $\bs{u}^{\rmsh}=\bs{u}^{out}$ at $r=b$) applied to Eq.~(\ref{eq:u_in}) and Eq.~(\ref{eq:u_out}):
\begin{equation}
    \begin{aligned}
     \text{At}\; r=a \quad : \quad &\frac{\rmd}{\rmd r}\left( \frac{h^{\rmsh}}{r} + \frac{\phi^{\rmsh}}{r}\right) =0, \\
     -&r \frac{\rmd}{\rmd r}\left(\frac{h^{\rmsh}}{r}\right) - 2 \frac{h^{\rmsh}}{r} + \frac{\phi^{\rmsh}}{r} = 1.\\
     \text{At}\; r=b \quad : \quad &\frac{\rmd}{\rmd r}\left( \frac{h^{\rmsh}}{r} + \frac{\phi^{\rmsh}}{r}\right) = \frac{\rmd}{\rmd r}\left( \frac{h^{\rmout}}{r} + \frac{\phi^{\rmout}}{r}\right), \\
     -&r \frac{\rmd}{\rmd r}\left(\frac{h^{\rmsh}}{r}\right) - 2 \frac{h^{\rmsh}}{r} + \frac{\phi^{\rmsh}}{r} = -r \frac{\rmd}{\rmd r}\left(\frac{h^{\rmout}}{r}\right) - 2 \frac{h^{\rmout}}{r} + \frac{\phi^{\rmout}}{r}.
    \end{aligned}
\end{equation}
In the above equation, we have used that there is no term with $\bs{x}$, when applying the condition $\bs u = \bs u^0$ in Eq.~(\ref{eq:u_in}). Now we can further use the identities: 
\begin{equation}
\begin{aligned}
\frac{\rmd}{\rmd r}\left(\frac{h}{r}\right) = -C\frac{a}{r^2}[1 + 3 G(k_T r)] \exp( \rmi k_T r) =-C \frac{a}{r^2}F(k_Tr)\exp(\rmi k_Tr),\\ \frac{\rmd}{\rmd r}\left(\frac{\phi}{r}\right) = -C\frac{a}{r^2}[1 + 3 G(k_L r)] \exp( \rmi k_L r)= -C\frac{a}{r^2}F(k_Lr) \exp(\rmi k_Lr)
\end{aligned}
\end{equation}
with $F(x)=-1-3\rmi/x+3/x^2=-1-3G(x)$, for the internal emitted ($h=h_e$, $\phi=\phi_e$) and transmitted external ($h=h_t$, $\phi=\phi_t$) waves and the same formulas with $G^*$ and $\exp(-\rmi kr)$ for the reflected formulas ($h=h_r$ and $\phi=\phi_r$). 

The stress continuity condition is more difficult to deduce in terms of $h$ and $\phi$. We need to calculate the stress tensor $\sigma_{ij}$ first with help of the displacement tensor $\epsilon_{ij}$ as\footnote{Note that $2\mu[ k_T^2/(2k_L^2)-1]=\lambda$, thus $\sigma_{ij}=\lambda \epsilon_{ll} \delta_{ij} + 2 \mu \epsilon_{ij}$.}:
\begin{equation}
    \begin{aligned}
     \frac{\sigma_{ij}}{2 \mu} &=\left( \frac{1}{2} \frac{k_T^2}{k_L^2}-1 \right) \epsilon_{ll} \delta_{ij} +\epsilon_{ij}\quad \quad ; \quad \quad
     \epsilon_{ij}&=\frac{1}{2}\left(\frac{\partial u_i}{\partial x_j} + \frac{\partial u_j}{\partial x_i} \right).
    \end{aligned}
\end{equation}
With Eq.~(\ref{eq:u_in}) or Eq.~\ref{eq:u_out}) we can get for the displacement tensors for the  L and T components:
\begin{equation}
    \begin{aligned}
     \frac{\partial u_i^L}{\partial x_j} =& \frac{\partial}{\partial x_j} \frac{\partial}{\partial x_i}\left[ \phi \frac{x_m u_m^0}{r}\right] =\frac{\partial u_j^L}{\partial x_i} =\frac{\rmd^2}{\rmd r^2}\left(\frac{\phi}{r}\right) \frac{x_i x_j}{r^2} x_m u_m^0 \\& + \frac{\rmd}{\rmd r}\left(\frac{\phi}{r} \right) \left[\delta_{ij} \frac{x_m u_m^0}{r} + \frac{x_i}{r} \delta_{jm} u_m^0 - \frac{x_i x_m}{r^3} u_m^0 x_j + u_i^0 \frac{x_j}{r} \right].
    \end{aligned}
\end{equation}
Thus:
\begin{equation}
    \begin{aligned}
     \epsilon_{ij}^L
     =\frac{\rmd^2}{\rmd r^2}\left(\frac{\phi}{r}\right) \frac{x_i x_j}{r^2} x_m u_m^0 + \frac{\rmd}{\rmd r}\left(\frac{\phi}{r} \right) \left[\delta_{ij} \frac{x_m u_m^0}{r} - \frac{x_i x_j}{r^3} u_m^0 x_m +  \frac{x_i u_j^0 + u_i^0x_j}{r} \right]
    \end{aligned}
\end{equation}
and $\epsilon_{ll}^L=\frac{\partial}{\partial x_l} \frac{\partial}{\partial x_l}\left(\phi x_k u_k^0/r  \right) = \nabla^2\left(\phi x_k u_k^0/r  \right)=\nabla^2\left(\phi \cos \theta  \right) u^0=-k_L^2 \boldsymbol x \cdot \boldsymbol u^0 \phi/r$. 

And similar for $\epsilon_{ij}^T$:
\begin{equation}
    \begin{aligned}
     \frac{\partial u_i^T}{\partial x_j} = &-2 u_i^0 \frac{\rmd}{\rmd r}\left(\frac{h}{r}\right)\frac{x_j}{r} +\frac{\rmd^2}{\rmd r^2}\left(\frac{h}{r}\right) \frac{x_j}{r}\left[x_i \frac{u_m^0 x_m}{r} -u_i^0 r \right] \\
     &+ \frac{\rmd}{\rmd r}\left(\frac{h}{r}\right)
    \left[ \delta_{ij} \frac{u_m^0 x_m}{r} + x_i \frac{u_m^0 \delta_{mj}}{r}- \frac{x_i x_j}{r^3} u_m^0 x_m -u_i^0 \frac{x_j}{r}\right],
    \end{aligned}
\end{equation}
thus
\begin{equation}
    \begin{aligned}
     \epsilon_{ij}^T =&-\frac{1}{2r}(u_i^0 x_j + u_j^0 x_i)3\frac{\rmd}{\rmd r}\left( \frac{h}{r}\right) + \frac{\rmd^2}{\rmd r^2}\left(\frac{h}{r}\right) \left[x_i x_j\frac{u_m^0 x_m}{r^2} -\frac{u_i^0 x_j}{2} - \frac{u_j^0 x_i}{2} \right] \\
     &+ \frac{\rmd}{\rmd r}\left(\frac{h}{r}\right)
    \left[ \delta_{ij} \frac{u_m^0 x_m}{r} - \frac{x_i x_j}{r^3} u_m^0 x_m +\frac{u_i^0 x_j}{2r} + \frac{u_j^0 x_i}{2r}\right] \\
    = & \frac{\rmd^2}{\rmd r^2}\left(\frac{h}{r}\right) \left[x_i x_j\frac{u_m^0 x_m}{r^2} -\frac{u_i^0 x_j}{2} - \frac{u_j^0 x_i}{2} \right] \\
     &+ \frac{\rmd}{\rmd r}\left(\frac{h}{r}\right)
    \left[ \delta_{ij} \frac{u_m^0 x_m}{r} - \frac{x_i x_j}{r^3} u_m^0 x_m -\frac{u_i^0 x_j}{r} - \frac{u_j^0 x_i}{r}\right]
    \end{aligned}
\end{equation}
and $\epsilon_{ll}^T=0$. For the traction we get
\begin{equation}
    \begin{aligned}
     \frac{f_i}{2 \mu}=\frac{\sigma_{ij}}{2 \mu}n_j = & x_i \frac{\boldsymbol x \cdot \boldsymbol u^0}{r^2}\left\{ - \left(\frac{1}{2}\frac{k_T^2}{k_L^2}-1\right)k_L^2\phi + \frac{r}{2}\frac{\rmd^2}{\rmd r^2}\left(\frac{h}{r} \right)\right .  \\&\quad \quad \quad \quad
     \left . - \frac{\rmd}{\rmd r}\left(\frac{h}{r} \right) + r \frac{\rmd^2}{\rmd r^2} \left( \frac{\phi}{r}\right) + \frac{\rmd}{\rmd r} \left( \frac{\phi}{r} \right) \right\}\\
    & + u_i^0 \left\{\frac{r}{2}\frac{\rmd^2}{\rmd r^2}\left( \frac{h}{r}\right) - \frac{\rmd}{\rmd r}\left( \frac{h}{r}\right) + \frac{\rmd}{\rmd r}\left( \frac{\phi}{r} \right) \right\}
    \end{aligned}
\end{equation}
where we have used $n_i = x_i/r$. For the second derivatives of $h$ and $\phi$ we get:
\begin{equation}
    \begin{aligned}
     \frac{\rmd^2}{\rmd r^2}\left(\frac{\phi}{r}\right) &= -C \frac{a}{r^3}\left[ -\rmi k_Lr + 5 +\frac{12 \rmi}{k_Lr} - \frac{12}{k_L^2 r^2}\right] \exp(\rmi k_L r)\\
     &=C \frac{a}{r^3}\left[k_L^2 r^2 G(k_Lr) +4 F(k_Lr)\right] \exp(\rmi k_Lr)
    \end{aligned}
\end{equation}
and similar for $h$ (replace $k_L$ by $k_T$) and the complex conjugates (replace $\rmi$ by $-\rmi$ everywhere).
Again we match the $x_i$ and $u_i^0$ terms of the traction to get
\begin{equation}
    \begin{aligned}  \label{eq:traction_xi}
    \frac{\mu^{\rmsh}}{\mu^{\rmout}} &\left\{ -  A^{\rmout}\phi_t + \frac{r}{2}\frac{\rmd^2}{\rmd r^2}\left(\frac{h_t}{r} \right)  - \frac{\rmd}{\rmd r}\left(\frac{h_t}{r} \right) + r \frac{\rmd^2}{\rmd r^2} \left( \frac{\phi_t}{r}\right) + \frac{\rmd}{\rmd r} \left( \frac{\phi_t}{r} \right) \right\}\\
   = & \; \; \; - A^{\rmsh}\phi_e + \frac{r}{2}\frac{\rmd^2}{\rmd r^2}\left(\frac{h_e}{r} \right)  - \frac{\rmd}{\rmd r}\left(\frac{h_e}{r} \right) + r \frac{\rmd^2}{\rmd r^2} \left( \frac{\phi_e}{r}\right) + \frac{\rmd}{\rmd r} \left( \frac{\phi_e}{r} \right)\\
    & \; \; \; - A^{\rmsh}\phi_r + \frac{r}{2}\frac{\rmd^2}{\rmd r^2}\left(\frac{h_r}{r} \right)  - \frac{\rmd}{\rmd r}\left(\frac{h_r}{r} \right) + r \frac{\rmd^2}{\rmd r^2} \left( \frac{\phi_r}{r}\right) + \frac{\rmd}{\rmd r} \left( \frac{\phi_r}{r} \right),
    \end{aligned}
\end{equation}
(with $A^{\rmout}=\left( \frac{1}{2}\frac{(k_T^{\rmout})^2}{(k_L^{\rmout})^2}-1\right)(k_L^{\rm\rmout})^2$ and $A^{\rmsh}=\left(\frac{1}{2}\frac{(k_T^{\rmsh})^2}{(k_L^{\rmsh})^2}-1\right)(k_L^{\rmsh})^2$) and
\begin{equation}
    \begin{aligned} \label{eq:traction_ui}
     \frac{\mu^{\rmsh}}{\mu^{\rmout}}&\left\{\frac{r}{2}\frac{\rmd^2}{\rm dr^2}\left( \frac{h_t}{r}\right) - \frac{\rmd}{\rmd r}\left( \frac{h_t}{r}\right) + \frac{\rmd}{\rmd r}\left( \frac{\phi_t}{r} \right) \right\}\\
     =& \;\; \; \; \frac{r}{2}\frac{\rmd^2}{\rmd r^2}\left( \frac{h_e}{r}\right) - \frac{\rmd}{ \rmd r}\left( \frac{h_e}{r}\right) + \frac{\rmd}{\rmd r}\left( \frac{\phi_e}{r}\right)\\
     &+\frac{r}{2}\frac{\rmd^2}{\rmd r^2}\left( \frac{h_r}{r}\right) - \frac{\rmd}{ \rmd r}\left( \frac{h_r}{r}\right) + \frac{\rmd}{\rmd r}\left( \frac{\phi_r}{r} \right).
    \end{aligned}
\end{equation}
The ratio $\mu^{\rmsh}/\mu^{\rmout}=(k^{\rmsh}_T/k^{\rmout}_T)^2 \rho_{\rmout}/\rho_{\rmsh}$ appears here (this is where the density ratio comes in). Eq.~(\ref{eq:traction_xi}) can be simplified by subtracting Eq.~(\ref{eq:traction_ui}) from it to get:
\begin{equation}
    \begin{aligned} \label{eq:traction_xi2}
    \frac{\mu^{\rmsh}}{\mu^{\rmout}} &\left\{ -  \left( \frac{1}{2}\frac{(k_T^{\rmout})^2}{(k_L^{\rmout})^2}-1\right)(k_L^{\rmout})^2\phi_t + r\frac{\rmd^2}{\rmd r^2}\left(\frac{h_t}{r} \right)   + r \frac{\rmd^2}{\rmd r^2} \left( \frac{\phi_t}{r}\right)  \right\}\\
   = & \; \; \;- \left(\frac{1}{2}\frac{(k_T^{\rmsh})^2}{(k_L^{\rmsh})^2}-1\right)(k_L^{\rmsh})^2\phi_e + r\frac{\rmd^2}{\rmd r^2}\left(\frac{h_e}{r} \right)  + r \frac{\rmd^2}{\rmd r^2} \left( \frac{\phi_e}{r}\right) \\
    & \; \; \; - \left(\frac{1}{2}\frac{(k_T^{\rmsh})^2}{(k_L^{\rmsh})^2}-1\right)(k_L^{\rmsh})^2\phi_r + r\frac{\rmd^2}{\rmd r^2}\left(\frac{h_r}{r} \right)  + r \frac{\rmd^2}{ \rmd r^2} \left( \frac{\phi_r}{r}\right). 
    \end{aligned}
\end{equation}
Eqs.~(\ref{eq:traction_ui}) and (\ref{eq:traction_xi2}) will give us 2 equations relating the $C_e^T$, $C_r^T$, $C_e^L$, $C_r^L$, $C_t^T$ and $C_t^L$. So the 2 traction equations, the 2 displacement equations at $r=a$ and the 2 displacement equations at $r=b$ will give us a $6 \times 6$ matrix system $A\cdot x=b$, with $x$ a vector containing the 6 unknown $C$'s.

\section{The solution in terms of spherical coordinates} \label{App:AppB}

From Eq.~(\ref{eq:u_in}) and (\ref{eq:u_out}), using the spherical coordinate system, we can predict that the solutions in the external domain and within the shell are, respectively,
\begin{equation}
\begin{aligned}
    \bs{u}^{\rmout} & = \bs{u}_{L}^{\rmout} + \bs{u}_{T}^{\rmout} \\&= \nabla [ U^0 a c_1 h^{(1)}_1(k^{\rmout}_{L}r)\cos{\theta} ] + \nabla \times [U^0 a c_2 h^{(1)}_1(k^{\rmout}_{T}r)\sin{\theta} \bs{e}_{\psi}],
\end{aligned}
\end{equation}
\begin{align}
    \bs{u}^{\rmsh}  = \bs{u}_{L}^{\rmsh} + \bs{u}_{T}^{\rmsh} \nonumber 
    = &\quad  \nabla \{ [U^0 a c_3 j_1(k^{\rmsh}_{L}r) + U^0 a c_4 y_1(k^{\rmsh}_{L}r)]\cos{\theta} \} \nonumber \\
    & + \nabla \times \{ [U^0 a c_5 j_1(k^{\rmsh}_{T}r) + U^0 a c_6 y_1(k^{\rmsh}_{T}r)]\sin{\theta} \bs{e}_{\psi} \}.
\end{align}
where $c_1 ... c_6$ are the unknown coefficients to be determined by the boundary conditions. Note that we have used the spherical Bessel functions of the first kind and the second kind, $j_{l}$ and $y_{l}$ ($l=1,2,3...$), for the solution in the shell and the spherical Hankel function of the first kind, $h^{(1)}_l$ ($l=1,2,3...$), for the solution in the external domain. The components of the displacement in the external domain are
\begin{subequations}
    \begin{align}
        u_{r}^{\rmout} &= U^0 a c_1 \frac{\del}{\del r}[h^{(1)}_1(k^{\rmout}_{L}r) \cos{\theta}] + U^0 a c_2 \frac{1}{r \sin{\theta}} \frac{\del}{\del r} \left[h^{(1)}_1(k^{\rmout}_{T}r)\sin{\theta}\right] \nonumber \\
        & =  U^0 a c_1 \frac{1}{r}[h^{(1)}_1(k^{\rmout}_{L}r) - k^{\rmout}_{L} r h^{(1)}_2(k^{\rmout}_{L}r)]\cos{\theta} + U^0 a c_2 \frac{2}{r}h^{(1)}_1(k^{\rmout}_{T}r) \cos{\theta},
    \end{align}
    \begin{align}
        u_{\theta}^{\rmout} &= U^0 a c_1 \frac{1}{r} \frac{\del}{\del \theta}[h^{(1)}_1(k^{\rmout}_{L}r) \cos{\theta}] - U^0 a c_2 \frac{1}{r} \frac{\del}{\del r} \left[r h^{(1)}_1(k^{\rmout}_{T}r)\sin{\theta}\right] \nonumber \\
        & = -U^0 a c_1 \frac{1}{r}h^{(1)}_1(k^{\rmout}_{L}r)\sin{\theta} - U^0 a c_2 \frac{1}{r} [2h^{(1)}_1(k^{\rmout}_{T}r) - k^{\rmout}_{T}r h^{(1)}_2(k^{\rmout}_{T}r) ]\sin{\theta},
    \end{align}
\end{subequations}
and $u^{\rmout}_{\psi} = 0$.

The components of the displacement in the shell are
\begin{subequations}
    \begin{align}
        u_{r}^{\rmsh} = & \quad U^0 a c_3 \frac{1}{r}[j_1(k^{\rmsh}_{L}r) - k^{\rmsh}_{L} r j_2(k^{\rmsh}_{L}r)]\cos{\theta} \nonumber\\
        & + U^0 a c_4 \frac{1}{r}[y_1(k^{\rmsh}_{L}r) - k^{\rmsh}_{L} r y_2(k^{\rmsh}_{L}r)]\cos{\theta} \nonumber \\
        & + U^0 a c_5 \frac{2}{r}j_1(k^{\rmsh}_{T}r) \cos{\theta} + U^0 a c_6 \frac{2}{r}y_1(k^{\rmsh}_{T}r) \cos{\theta},
    \end{align}
    \begin{align}
        u_{\theta}^{\rmsh} = & -U^0 a c_3 \frac{1}{r}j_1(k^{\rmsh}_{L}r)\sin{\theta} -U^0 a c_4 \frac{1}{r}y_1(k^{\rmsh}_{L}r)\sin{\theta} \nonumber \\        
        &- U^0 a c_5 \frac{1}{r} [2j_1(k^{\rmsh}_{T}r) - k^{\rmsh}_{T}r j_2(k_{T}r) ]\sin{\theta} \nonumber \\ 
        &- U^0 a c_6 \frac{1}{r} [2y_1(k^{\rmsh}_{T}r) - k^{\rmsh}_{T}r y_2(k_{T}r) ]\sin{\theta},
    \end{align}
\end{subequations}
and $u^{\rmsh}_{\psi} = 0$.

In the shell-external coupling part, we need to match the traction across the interface as well. In spherical coordinate system, we have~\citep{Hinders1991}
\begin{subequations}
    \begin{align}
        \sigma_{rr} = &  \lambda \nabla \cdot \bs{u} + 2 \mu \frac{\del u_r}{\del r} \nonumber \\
        = & \lambda \left[ \frac{1}{r^2}\frac{\del}{\del r}\left(r^2 u_r\right) + \frac{1}{r \sin{\theta}} \frac{\del}{\del\theta} (\sin{\theta}u_{\theta}) + \frac{1}{r\sin{\theta}} \frac{\del u_{\psi}}{\del \psi} \right] + 2 \mu \frac{\del u_r}{\del r}, \\
        \sigma_{r\theta} = & \mu \left( \frac{\del u_{\theta}}{\del r} -\frac{u_{\theta}}{r} +  \frac{1}{r} \frac{\del u_{r}}{\del \theta} \right), \\
        \sigma_{r\psi} = & \mu \left( \frac{\del u_{\psi}}{\del r} -\frac{u_{\psi}}{r} +  \frac{1}{r \sin{\theta}} \frac{\del u_{r}}{\del \psi} \right)=0.
    \end{align}
\end{subequations}
In the above equations, we only need to match $\sigma_{rr}$ and $\sigma_{r\theta}$ across the interface since $\sigma_{r\psi}$ is zero due to the symmetry considerations: $u_\psi=0$ and $\partial/\partial \psi = 0$. We then have, in the external domain,
\begin{subequations}
    \begin{align}
        \sigma_{rr}^{\rmout} = &- \lambda^{\rmout} U^0 a c_1 (k_{L}^{\rmout})^2 h^{(1)}_{1}(k_{L}^{\rmout} r) \cos{\theta} \nonumber \\ 
        & + \mu^{\rmout} U^0 a c_1 \left[ \frac{4 k_{L}^{\rmout}}{r} h^{(1)}_{2}(k_{L}^{\rmout} r) -2 (k_{L}^{\rmout})^2 h^{(1)}_{1}(k_{L}^{\rmout} r) \right] \cos{\theta} \nonumber \\
        & - \mu^{\rmout} U^0 a c_2 \frac{4 k_{T}^{\rmout}}{r} h^{(1)}_{2}(k_{T}^{\rmout} r) \cos{\theta}, \\
        \sigma_{r\theta}^{\rmout} = & \quad \mu^{\rmout} U^0 a c_1 \frac{2k_{L}^{\rmout}}{r} h^{(1)}_{2}(k_{L}^{\rmout} r) \sin{\theta} \nonumber \\
        & -\mu^{\rmout} U^0 a c_2 \left[ \frac{2 k_{T}^{\rmout}}{r} h^{(1)}_{2}(k_{T}^{\rmout} r) - (k_{T}^{\rmout})^2 h^{(1)}_{1}(k_{T}^{\rmout} r) \right] \sin{\theta}.
    \end{align}
\end{subequations}
In the shell, 
\begin{subequations}
    \begin{align}
        &\sigma_{rr}^{\rmsh} = - \lambda^{\rmsh} U^0 a c_3 (k_{L}^{\rmsh})^2 j_{1}(k_{L}^{\rmsh} r) \cos{\theta} - \lambda^{\rmsh} U^0 a c_4 (k_{L}^{\rmsh})^2 y_{1}(k_{L}^{\rmsh} r) \cos{\theta} \nonumber \\ 
        & + \mu^{\rmsh} \left\{ U^0 a c_3 \left[ \frac{4 k_{L}^{\rmsh}}{r} j_{2}(k_{L}^{\rmsh} r) -2 (k_{L}^{\rmsh})^2 j_{1}(k_{L}^{\rmsh} r) \right] - U^0 a c_5 \frac{4 k_{T}^{\rmsh}}{r} j_{2}(k_{T}^{\rmsh} r)  \right\}\cos{\theta} \nonumber \\ 
        & + \mu^{\rmsh} \left\{ U^0 a c_4 \left[ \frac{4 k_{L}^{\rmsh}}{r} y_{2}(k_{L}^{\rmsh} r) -2 (k_{L}^{\rmsh})^2 y_{1}(k_{L}^{\rmsh} r) \right] - U^0 a c_6 \frac{4 k_{T}^{\rmsh}}{r} y_{2}(k_{T}^{\rmsh} r)  \right\}\cos{\theta}, \\
        &\sigma_{r\theta}^{\rmsh} = \mu^{\rmsh} U^0 a c_3 \frac{2k_{L}^{\rmsh}}{r} j_{2}(k_{L}^{\rmsh} r) \sin{\theta} - \mu^{\rmsh}U^0 a c_5 \left[ \frac{2 k_{T}^{\rmsh}}{r} j_{2}(k_{T}^{\rmsh} r) - (k_{T}^{\rmsh})^2 j_{1}(k_{T}^{\rmsh} r) \right] \sin{\theta} \nonumber \\
        & +  \mu^{\rmsh}U^0 a c_4 \frac{2k_{L}^{\rmsh}}{r} y_{2}(k_{L}^{\rmsh} r) \sin{\theta} - \mu^{\rmsh}U^0 a c_6 \left[ \frac{2 k_{T}^{\rmsh}}{r} y_{2}(k_{T}^{\rmsh} r) - (k_{T}^{\rmsh})^2 y_{1}(k_{T}^{\rmsh} r) \right] \sin{\theta}.
    \end{align}
\end{subequations}
By matching the two displacement components $u_r$, $u_{\theta}$ on the vibrating core sphere and the interface between the shell and the external domain as well as the normal and tangential stress components $\sigma_{rr}$, $\sigma_{r\theta}$ across the shell boundary, we obtain a $6 \times 6$ linear system to solve for the unknown coefficients $c_1$ to $c_6$.

\section{Validation of a boundary element method}\label{App:BEM}
\begin{figure*}[t]  
\begin{center} 
\includegraphics[width=0.75\textwidth]{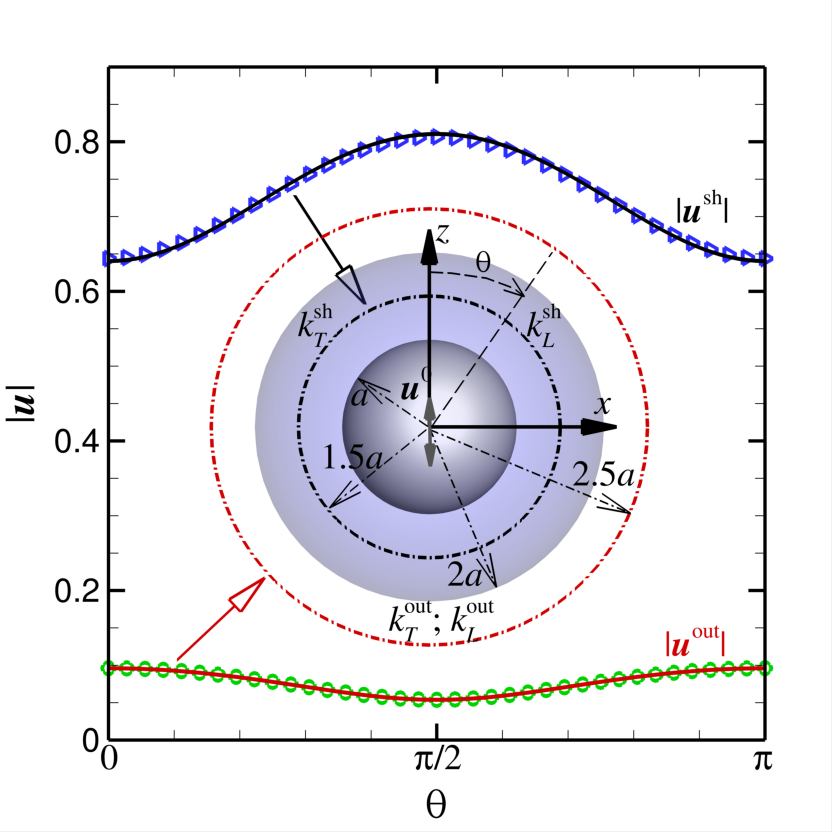}
    \caption{Excellent agreement is found for a comparison between the analytical solution (solid curves) and the numerical solution by BEM (symbols), $\rho_{\rmout}/\rho_{\rmsh} =3.0$, $b/a=2.0$, $\boldsymbol u^0 = (0,0,1)$, $k_T^{\rmsh} a =4.5$, $k_L^{\rmsh} a=2.0$, $k_T^{\rmout}a=2.0$ and $k_L^{\rmout} a=1.0$. Two concentric circles, one with radius $r=2.5a$ for the outer domain and another with radius $r=1.5a$ for the shell are used for the comparison (both situated on the $xz$ plane). $| \bs u|$ is used for the comparison.} \label{Fig:ValBEM}
\end{center}
\end{figure*}
The analytical solution developed in this work is an ideal tool to benchmark numerical tools since it is a simple yet non-trivial solution that contains both outgoing and reflected wave phenomena in linear elasticity. In this appendix, we will use the solution in Sec.~\ref{sec:coreshellsol} and~\ref{App:AppB} to validate a vector boundary element method reported in~\cite{Rizzo1985}. As shown in Fig.~\ref{Fig:ValBEM}. The method of Rizzo works with the total displacement vector $\bs u$ and is not capable of distinguishing between $\bs u_T$ and $\bs u_L$. Excellent agreement is observed between the numerical method and the analytical solution.




\bibliographystyle{elsarticle-harv}
\bibliography{sample}





\end{document}